\documentclass[print]{aa}
\usepackage[varg]{txfonts}
\usepackage{physics}
\usepackage{subcaption}
\usepackage{textcomp}
\usepackage{booktabs}
\usepackage{graphicx}
\usepackage{gensymb}
\usepackage{multirow}

\newcommand{\red}[1]{{\color{black}#1}}
\begin{document}

\title{Large Interferometer For Exoplanets (LIFE):}

\titlerunning{LIFE: VII. Beam combiner implementation, instrumental uncertainties and telescope redundancies}

\authorrunning{Hansen et al.}

\subtitle{VII. Practical implementation of a \red{five-telescope} kernel-nulling beam combiner with a discussion on instrumental uncertainties and redundancy benefits}

\author{Jonah T. Hansen\inst{1}\thanks{Correspondence: \href{mailto:jonah.hansen@anu.edu.au}{jonah.hansen@anu.edu.au}},
  Michael J. Ireland\inst{1}\and Romain Laugier\inst{2} \and the LIFE Collaboration\inst{3}}

\institute{Research School of Astronomy and Astrophysics, College of Science, Australian National University, Canberra, Australia, 2611
\and 
Institute of Astronomy, KU Leuven, Celestijnenlaan 200D, 3001, Leuven, Belgium \and
\url{www.life-space-mission.com}}

\date{Received X XXX XXXX}


\abstract {In the \red{fourth} paper in this series, we identified that a pentagonal arrangement of five telescopes, using a kernel-nulling beam combiner, shows notable advantages for some important performance metrics for a space-based mid-infrared nulling interferometer over several other considered configurations for the detection of Earth-like exoplanets around solar-type stars.} {We aim to produce a \red{physical} implementation of a kernel-nulling beam combiner for such a configuration, as well as a discussion of systematic and stochastic errors associated with the instrument.} {We developed a mathematical framework around a \red{nulling beam combiner}, and then used it along with a \red{space interferometry simulator} to identify the effects of systematic uncertainties.} {We find that errors in the beam combiner optics, systematic phase errors and the root-mean-squared (RMS) fringe tracking errors result in instrument-limited performance at $\sim$4-7~\textmu m, and zodiacal light limited at $\gtrsim$10~\textmu m. Assuming a beam splitter reflectance error of $|\Delta R| = 5\%$ and phase shift error of $\Delta\phi = 3\degree$, we find that the fringe tracking RMS error should be kept to less than 3~nm in order to be photon limited, and the systematic piston error be less than 0.5~nm to be appropriately sensitive to planets with a contrast of 1$\times 10^{-7}$ over a 4-19~\textmu m bandpass. We also identify that the beam combiner design, with the inclusion of a well-positioned shutter, provides an ability to produce robust kernel observables even if one or two collecting telescopes were to fail. The resulting four-telescope combiner, when put into an X-array formation, results in a transmission map with a relative signal-to-noise ratio equivalent to 80\% of a fully functioning X-array combiner.} {The advantage in sensitivity and planet yield of the Kernel-5 nulling architecture, along with an inbuilt contingency option for a failed collector telescope, leads us to recommend this architecture be adopted for further study for the LIFE mission.}

\maketitle 

\section{Introduction}
\label{sec:intro}
Optical/mid-infrared nulling interferometry from space has been experiencing a resurgence of interest over the past few years, particularly with regards to detecting Earth-like exoplanets around solar-type stars. Such an idea is not new, having been first proposed by Bracewell \citep{1978Bracewell}, and then through multiple studies resulting in two large missions: the European Space Agency's \textit{Darwin} \citep{Leger1996} and NASA's \textit{Terrestrial Planet Finder - Interferometer} (TPF-I) \citep{Beichman1999}. However, due to a myriad of reasons, not least concerning the lack of technological readiness, both missions were cancelled in the late 2000s.

Since then, various teams have continued to work \red{on improving} nulling interferometry, leading to the formation of the \textit{Large Interferometer For Exoplanets} (LIFE) initiative. This project is being considered as one of the large-class missions of the European Space Agency's Voyage 2050 programme \citep{2021ESAVoyage}: a large space interferometer in the legacy of \textit{Darwin}, working in the mid-infrared, with a goal to both detect and characterise Earth-like exoplanets that are difficult to access using other techniques such as single aperture coronography and transit spectroscopy. Significant work has already been done to characterise the planet yield of such a mission \citep{2018Kammerer,LIFE1}, and the spectral requirements of the instrument \citep{LIFE3}. A simulator tool to simulate observations and signal-to-noise ratio (S/N) requirements has also been developed \citep{LIFE2}.

The renewal of interest in space interferometry also presents an opportune time to reanalyse the technology behind nulling interferometry. In particular, new technologies such as `kernel-nulling' \citep{2018Martinache,Laugier2020} have opened up avenues to consider other telescope configurations away from the Emma X-array configuration decided upon in the \textit{Darwin}/TPF-I era \citep{2005LayXarray}. The \red{fourth} paper in this series \citep[hereby LIFE4]{LIFE4} conducted a trade study between a number of different configurations, including the X-array, to determine whether other architectures would provide a higher yield and higher S/N. It was found that, in fact, an architecture consisting of five telescopes in a pentagonal shape \citep[e.g.][]{Leger1996,1997Mennesson}, using a kernel-nulling beam combination scheme, outperformed the X-array in both detection and characterisation.

In this paper, we propose a practical way of implementing the \red{five-telescope} beam combination scheme discussed in the previous paper. We also discuss the systematic instrumental errors of such a beam combiner: how these change the dominant sources of photon noise, and how they impact the robustness and sensitivity of the kernel observable. Finally, we discuss a major advantage of this beam combination scheme---that even if a collector telescope is damaged or fails, the interferometer and beam combiner are still able to produce robust transmission maps with fewer telescopes.

\section{Implementation of the beam combination scheme}
\label{sec:implementation}
\begin{figure}
    \centering
    \includegraphics[width=\linewidth]{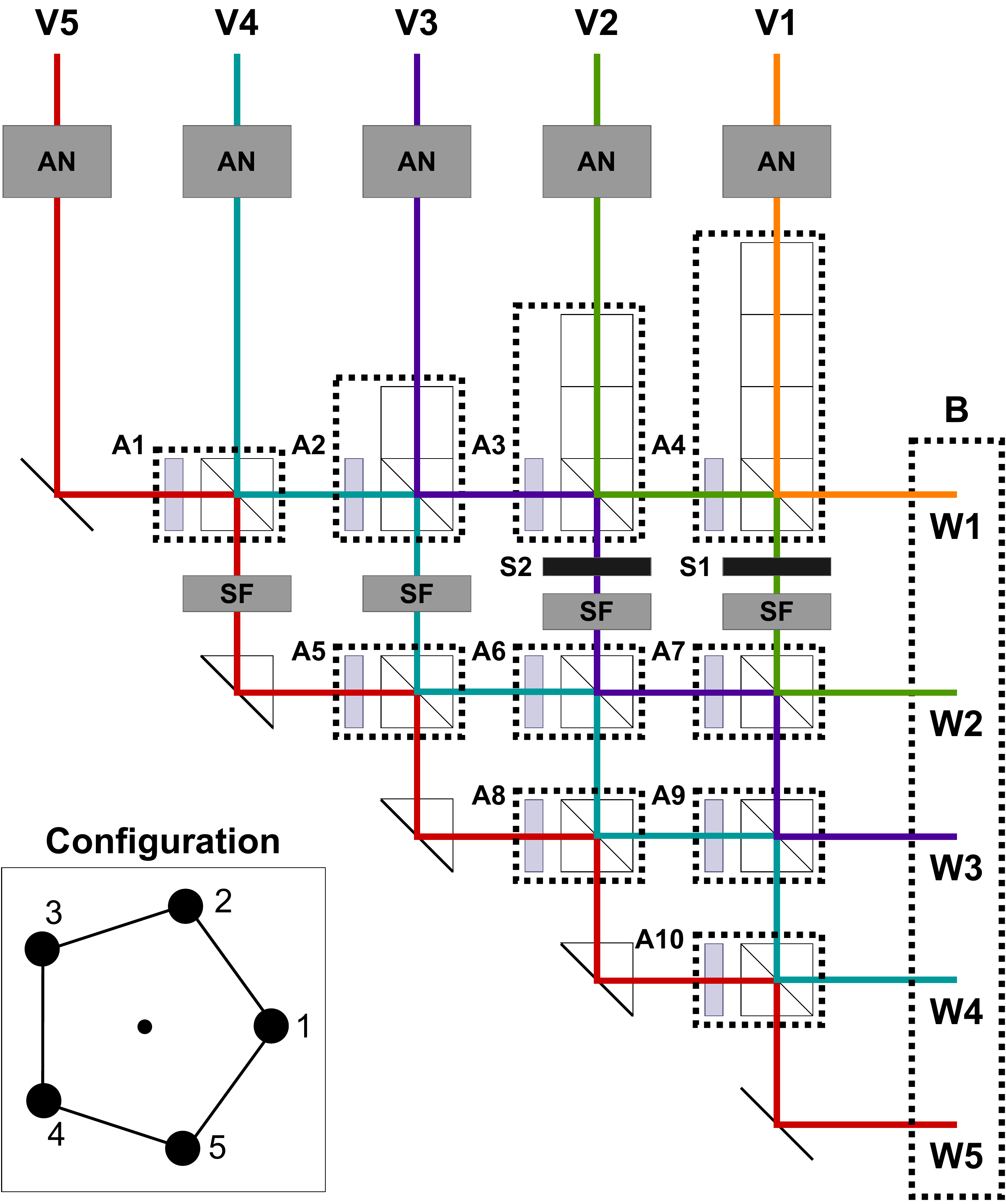}
    \caption{Schematic of a Kernel-5 beam combiner, based on the design of \cite{Guyon2013}. Inputs $\vb{V}_1$ through $\vb{V}_5$ pass through five adaptive nuller units ($\vb{AN}$), a series of ten beam splitter modules ($\vb{A}_1$ through $\vb{A}_{10}$) consisting of a beam splitter and phase shifting plate on one input, and four spatial filters ($\vb{SF}$). The five outputs consist of one bright output $\vb{W}_1$ and four nulled outputs $\vb{W}_2$ through $\vb{W}_4$. Two shutters ($\vb{S}$) can be used in case of a telescope failure (see Section \ref{sec:redundancy}). \red{The inset shows the telescope configuration and their corresponding inputs.}}
    \label{Img:kernel-5-guyon}
\end{figure}

We devised an implementation of the Kernel-5 nuller beam combiner through the method of \cite{Guyon2013}. They posit that any predetermined unitary lossless transfer matrix $\vb{M}$ (denoted $U$ in their notation) of $m$ inputs, can be created through a series of $n = \frac{m(m-1)}{2}$ unequal beam splitters, with a phase shifting plate put in front of one of the inputs of each beam splitter. Such a design for a five-telescope combiner, can be seen in Figure \ref{Img:kernel-5-guyon}. This design also includes a set of $m$ adaptive nullers (denoted $\vb{AN}$) and $m-1$ spatial filters (denoted $\vb{SF}$) that are used to remove systematic amplitude and phase errors on input. \red{We note here that the additional glass prisms on the top row of beam splitters are to ensure that the path lengths are matched \red{at all wavelengths by passing through} equal amounts of air and glass. We have also indicated that three of the fold mirrors should be total internal reflection (TIF) prisms or similar for this same reason. This design implicitly assumes that the path lengths of the beams are matched before entering the first prisms.}

\red{The design can be broadly broken down into two parts: the first row of beam splitters before the spatial filters perform the nulling, and the remaining beam splitters perform mixing in order to create kernel outputs.} The spatial filters are placed after the first row of beam splitters \red{so that a precise optical alignment is not required to get a deep null.} These will be discussed further in Section \ref{sec:uncertainties}. Finally, we also have included two shutters in the design (denoted $\vb{S}$); these shutters can be used in the case of a collector telescope failure to reconfigure the beam combiner to produce robust observables with fewer telescopes. This will be discussed in more detail in Section \ref{sec:redundancy}. The parameters for each beam splitter and phase shifter can be derived through working backwards from the predetermined matrix $\vb{M}$.

We start with the general case of a beam combiner with $m$ inputs (labelled $\vb{V}_1$ through $\vb{V}_m$) and $m$ outputs ($\vb{W}_1$ through $\vb{W}_m$) as depicted for $m=5$ in Figure \ref{Img:kernel-5-guyon}. A phase shifting plate is put in front of the second input of each beam splitter (the beam entering from the left in the figure), imposing a phase shift of $\phi_j$ for each $j$th plate. We define each beam splitter through a mixing angle $\theta$, which is related to their reflectance $R$ and transmittance $T$ as follows:
\begin{align}
    R = \sin{\theta} && T = \cos{\theta}.
\end{align}
Hence each phase shifter and beam splitter module can be described by a 2x2 matrix:
\begin{equation}
    \vb{C}_j = \begin{bmatrix}
    \sin{\theta_j} & e^{i\phi_j}\cos{\theta_j}\\
    \cos{\theta_j} & -e^{i\phi_j}\sin{\theta_j}
    \end{bmatrix}.
\end{equation}

As we have $m$ input beams, with only two interfering at any one time, the other $m-2$ beams are represented by identity rows and columns. Each beam combining step can thus be represented as an $m\times m$ block diagonal matrix $\vb{A}_j$, with the diagonals of the rows and columns corresponding to the combining beams being equal to $\vb{C}_j$ and having ones on the other diagonal terms. For example, for a beam splitter module combining beams two and three out of a five beam combiner, the matrix $\vb{A}_j$ is given by
\begin{equation}
    \vb{A}_j = \begin{bmatrix}
    1 & \vb{0}^T & 0 & 0\\
    \vb{0} & \vb{C}_j & \vb{0} & \vb{0}\\
    0 & \vb{0}^T & 1 & 0\\
    0 & \vb{0}^T & 0 & 1
    \end{bmatrix},
\end{equation}
where $\vb{0} = [0,0]$ is a two element zero column vector.

The full beam combiner is hence described by a multiplication of the $n$ $\vb{A}_j$ matrices. We also note that each of the output electric fields of the beam combiner can have an arbitrary phase shift ($\omega$) relative to the first output, as we only measure the intensity of these beams. We represent this as the matrix $\vb{B}$, given by
\begin{equation}
    \vb{B} = \begin{bmatrix}
    1 & 0 & ... & 0\\
    0 & e^{i\omega_2} & ... & 0\\
    ... & ... & ... & 0\\
    0 & 0 & 0 & e^{i\omega_m}\\
    \end{bmatrix}.
\end{equation}

Therefore, to create a beam combiner for any transfer matrix $\vb{M}$, we solve the following equation for parameters $\theta_j$, $\phi_j$ ($j = 1,..,n$) and $\omega_k$ ($k = 2,...,m$). This equation is built from the $\vb{A}_j$ matrices in ascending order, corresponding to the order in which light traverses the combiner from the top left corner to the bottom right:
\begin{equation}
\label{eq:solve}
    \vb{M} = \vb{B}\vb{A}_n\vb{A}_{n-1}...\vb{A}_2\vb{A}_1.
\end{equation}

For the Kernel-5 nuller there are $m=5$ inputs and thus $n=10$ beam splitters, and we know the transfer matrix $\vb{M}$ from LIFE4:
    \begin{equation}
   \vb{M} = \frac{1}{\sqrt{5}}\begin{bmatrix}
    1 & 1 & 1 & 1 & 1\\
    1 & e^{\frac{-4\pi i}{5}} & e^{\frac{2\pi i}{5}} & e^{\frac{-2\pi i}{5}} & e^{\frac{4\pi i}{5}}\\
    1 & e^{\frac{4\pi i}{5}} & e^{\frac{-2\pi i}{5}} & e^{\frac{2\pi i}{5}} & e^{\frac{-4\pi i}{5}}\\
    1 & e^{\frac{2\pi i}{5}} & e^{\frac{4\pi i}{5}} & e^{\frac{-4\pi i}{5}} & e^{\frac{-2\pi i}{5}}\\
    1 & e^{\frac{-2\pi i}{5}} & e^{\frac{-4\pi i}{5}} & e^{\frac{4\pi i}{5}} & e^{\frac{2\pi i}{5}}
\end{bmatrix}.
\end{equation}

\red{This combiner design will produce two sets of second-order nulls, forming one kernel, and two sets of fourth-order nulls, forming the other kernel. Here, a second-order null is one where the interferometer throughput scales as $|\alpha|^2$, $|\alpha|$ being the angular separation from the optical axis, and a fourth-order null scaling as $|\alpha|^4$. In this investigation, we have swapped rows two and four from the matrix in LIFE4 so that the two kernel outputs are formed from neighbouring pairs (with the second order kernel being $|\vb{W}_4|^2-|\vb{W}_5|^2$ and the fourth order being $|\vb{W}_2|^2-|\vb{W}_3|^2$).} Assuming the pentagonal formation described in LIFE4, this also allows us to form the deeper, \red{fourth-order} null, calculated from the difference in rows two and three, using fewer beam splitting modules.

\begin{figure}
  \centering
  \begin{subfigure}{0.8\linewidth}
    \centering
    \includegraphics[width=\linewidth]{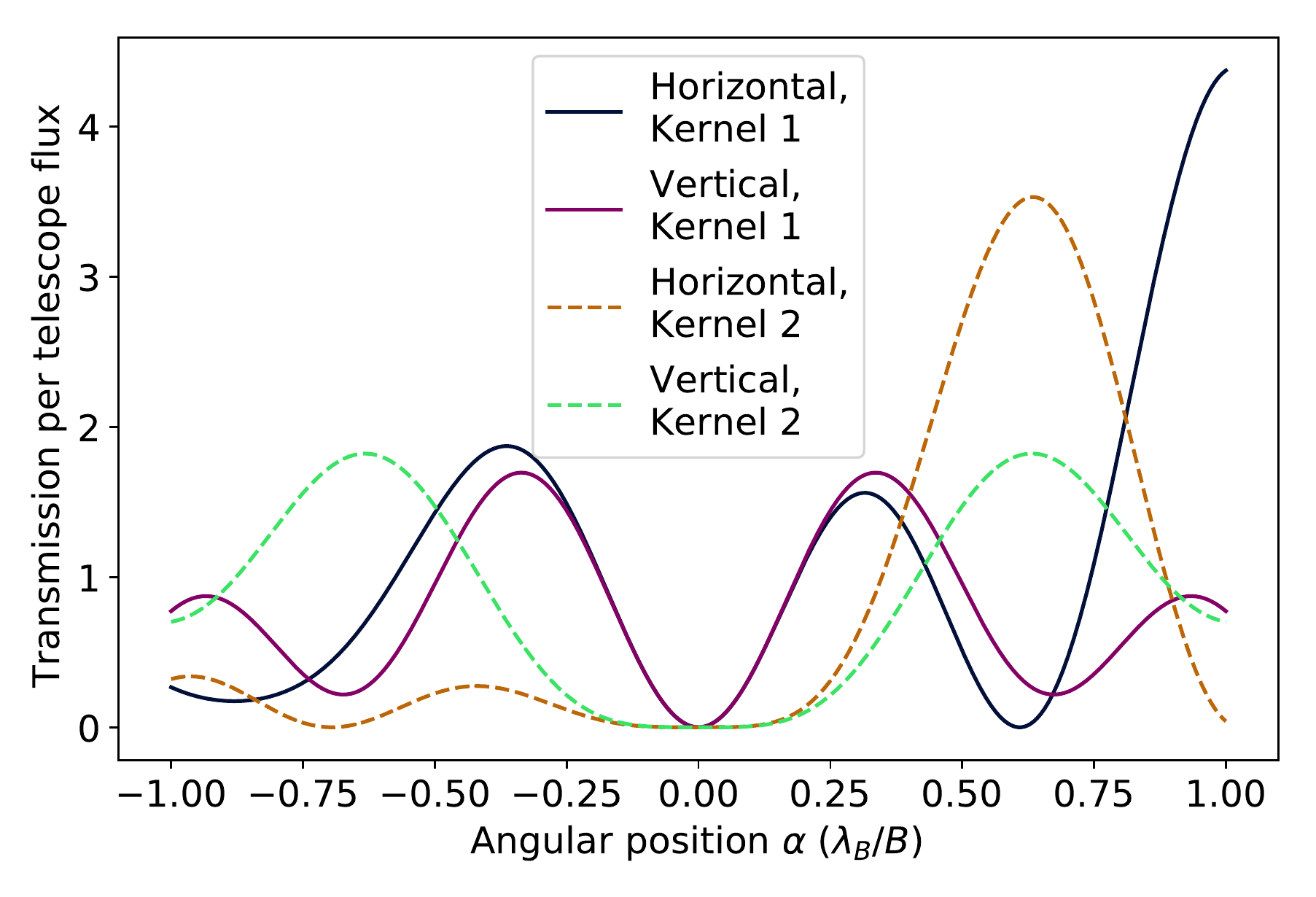}
   \caption{\red{Linear Scaling}}
   \label{Img:null_order_linear}
  \end{subfigure}
  \hfill
  \begin{subfigure}{0.8\linewidth}
    \centering
    \includegraphics[width=\linewidth]{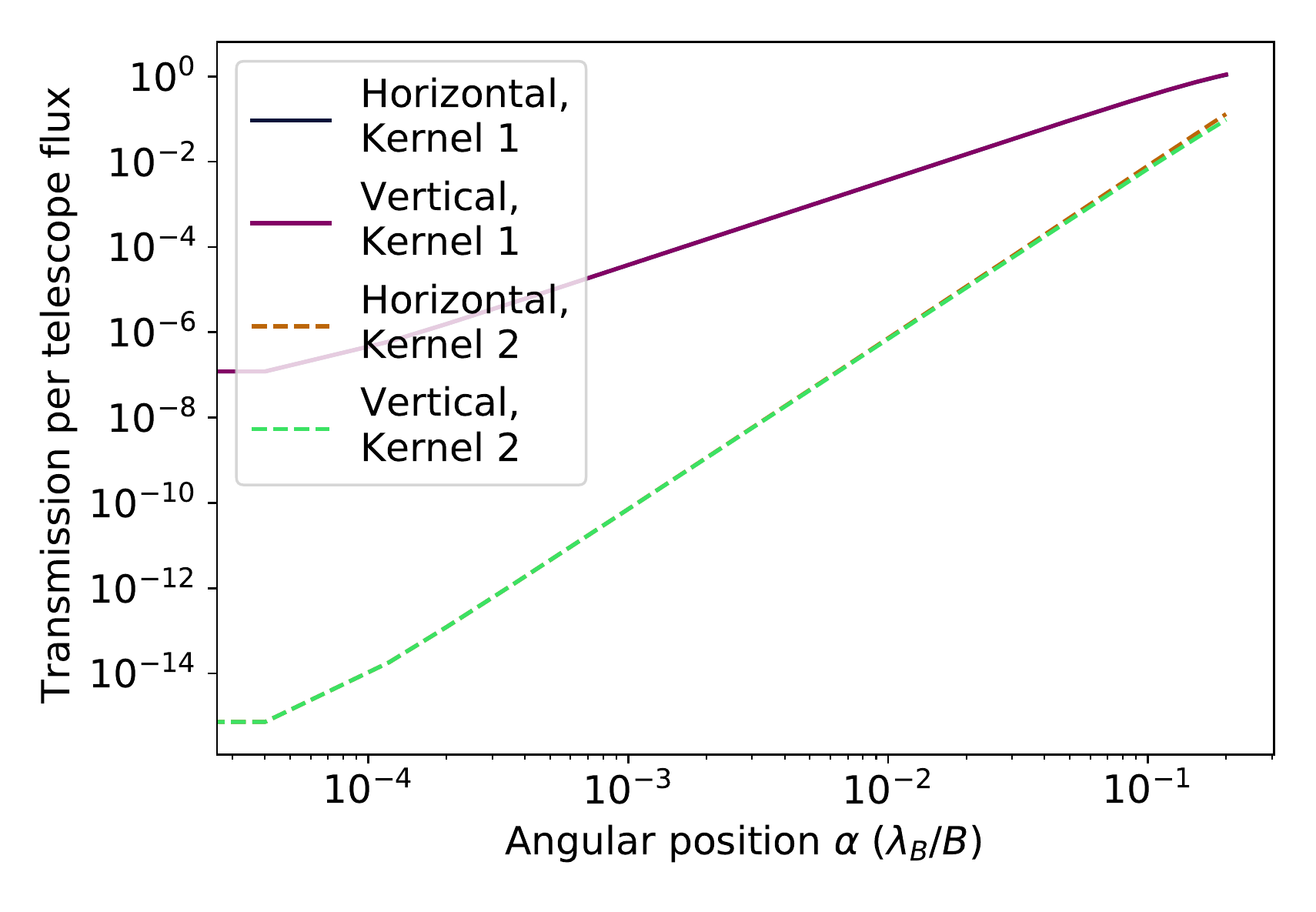}
   \caption{\red{Log-Log scaling}}
   \label{Img:null_order_log}
  \end{subfigure}
    \caption{\red{Instrument transmission per telescope flux as a function of angular separation from the optical axis for each output pair corresponding to the two kernels (i.e. $\vb{W}_4$ for kernel 1 and $\vb{W}_2$ for kernel 2). For each output, a slice through the optical axis was taken in the horizontal and vertical directions; due to the symmetry about the horizontal axis shown in the inset in Figure \ref{Img:kernel-5-guyon}, the vertical component is symmetric about zero.}}
    \label{Img:null_orders}
\end{figure}

To be consistent between the naming schemes of the this paper and LIFE4, we define kernel 1 as the second-order null and kernel 2 as the fourth-order null. \red{We plot the instrument transmission for the outputs corresponding to each kernel as a function of angular position in Figure \ref{Img:null_orders}, in both a linear (\ref{Img:null_order_linear}) and log-log (\ref{Img:null_order_log}) scaling. We have plotted this transmission in both the horizontal and vertical directions, assuming the configuration shown in the inset of Figure \ref{Img:kernel-5-guyon}; as such the vertical slice is symmetric about zero. Here, we can clearly see that kernel 2 has a deeper, broader null and is thus less affected by stellar leakage in comparison to kernel 1.} \red{Furthermore, from the slopes of the lines in Figure \ref{Img:null_order_log}, we confirm the second and fourth order relations of the nulls with angular position, and we can generate a mathematical description of the nulls, being
\begin{align}
    |W_4|^2 = |W_5|^2 \approx 37.2\left(\frac{|\alpha|B}{\lambda}\right)^2\\
    |W_2|^2 = |W_3|^2 \approx 70.8\left(\frac{|\alpha|B}{\lambda}\right)^4.
\end{align}}

Working through element by element, we solve Equation \ref{eq:solve} for this matrix $\vb{M}$, resulting in the parameters listed in Table \ref{tab:kernel-5_params}. The output phase shifts $\omega$ are all zero except for $\omega_5$, which has a phase shift of $\pi$. \red{We note here that the phase shifts listed do not account for the phase shifts induced by the fold mirrors and TIF prisms. However, these can be compensated for by adding the additional phase shift (nominally $\pi$ for a simple mirror reflection) before the relevant beam splitters (i.e. modifying $\phi_1$, $\phi_5$, $\phi_8$ and $\phi_{10}$) and as this should be a constant offset, it will not interfere with the following analysis. }

\begin{table*}[]
\centering
\caption{Optical parameters for the beam combiner design discussed in Section \ref{sec:implementation} and displayed in Figure \ref{Img:kernel-5-guyon}.}
\label{tab:kernel-5_params}
\begin{tabular}{@{}cccc@{}}
\toprule
              & Mixing Angle ($\theta$)                            & Reflectance coefficient $|R|$              & Phase Shift ($\phi$)                                  \\ \midrule
$\vb{C}_1$    & -$\frac{\pi}{4}\approx-0.785$                                   & $\frac{1}{\sqrt{2}} \approx 0.707$                     & $\pi$                                                 \\
$\vb{C}_2$    & $\arcsin{\left(\frac{1}{\sqrt{3}}\right)}\approx0.615$                      & $\frac{1}{\sqrt{3}} \approx 0.577$                     & $\pi$                                                 \\
$\vb{C}_3$    & $\frac{5\pi}{6}\approx2.618$                                   & $\frac{1}{2}$                            & $\pi$                                                 \\
$\vb{C}_4$    & $\pi-\arcsin{\left(\frac{1}{\sqrt{5}}\right)}\approx2.678$                  & $\frac{1}{\sqrt{5}}\approx0.447$                     & $\pi$                                                 \\
$\vb{C}_5$    & $\arctan{\left(\sqrt{\frac{1}{3}(4-\sqrt{5})}\right)}\approx0.654$          & $\sqrt{\frac{4-\sqrt{5}}{7-\sqrt{5}}} \approx 0.608$  & $-\frac{3\pi}{10}-\arctan{\left(\sqrt{5-\frac{2}{\sqrt{5}}}\right)}\approx-2.055$ \\
$\vb{C}_6$    & $\pi-\arcsin{\left(\frac{1}{3}\sqrt{3-\frac{2}{\sqrt{5}}}\right)}\approx2.637$ & $\frac{1}{3}\sqrt{3-\frac{2}{\sqrt{5}}} \approx 0.484$ & $ \arctan{\left(\frac{3}{7}\sqrt{5-2\sqrt{5}}\right)}-\pi\approx-2.840$              \\
$\vb{C}_7$    & $-\frac{\pi}{6}\approx-0.524$                                   & $\frac{1}{2}$                            & $\arctan{\left(\sqrt{2-\frac{2}{\sqrt{5}}}\right)}\approx0.810$                \\
$\vb{C}_8$    & $\arcsin{\left(\frac{-1+3\sqrt{5}}{2\sqrt{22}}\right)}-\pi\approx-2.487$    & $\frac{-1+3\sqrt{5}}{2\sqrt{22}}\approx 0.608$  & $\pi-\arctan{\left(\frac{1}{\sqrt{25+10\sqrt{5}}}\right)}\approx2.997$           \\
$\vb{C}_9$    & $-\arcsin{\left(\frac{1}{\sqrt{3}}\right)}\approx-0.615$                     & $\frac{1}{\sqrt{3}} \approx 0.577$                     & $-\arctan{\left(\sqrt{\frac{1}{10}(5-\sqrt{5})}\right)}\approx-0.484$               \\
$\vb{C}_{10}$ & $-\frac{\pi}{4}\approx-0.785$                                   & $\frac{1}{\sqrt{2}} \approx 0.707$                     & -$\frac{\pi}{2}\approx-1.571$                                     \\ \bottomrule
\end{tabular}%
\end{table*}

We note here \red{several remarks on} this implementation of the beam combiner. Firstly, the beam splitters and phase shifting plates are required to \red{have achromatic phase shifts and reflection coefficients (with tolerances discussed in Section~\ref{sec:opticalerrors})} over a large wavelength band (nominally $4-19$~\textmu m \citep{LIFE1}). This could be alleviated by increasing the number of beam trains, splitting the wavelength into a few coarse channels, and implementing multiple versions of the beam combiner. The downside to this method is the increase of optical components and space requirements; a pertinent problem for a space-based mission. We return to this issue when we discuss phase chopping in Section \ref{sec:phase_chop}.
    
Secondly, this design is inherently polarisation dependent. For this reason, we notionally assume that the beam combiner is planar and there is a polarisation split orthogonal to the plane before the telescope light is injected into the combiner unit. Thus we do not explicitly consider polarisation effects in the following discussions.
    
Thirdly, we have included the spatial filters after the nulling stage of the beam combiners so that aberrations in this first stage can be compensated with upstream corrective optics or the adaptive nullers as described in Section \ref{sec:uncertainties}. If spatial filtering occurred before the nulling stage, then any alignment errors or optical aberrations in the beamsplitters could not be corrected and would result in a decreased null depth. 
    
Finally, this implementation is schematically drawn as bulk optics. While photonics provides multiple advantages in terms of spatial filtering and space requirements (in terms of both footprint and space compatibility), adequate demonstrations of far mid-infrared achromatic directional couplers and phase shifts have not, to the authors' knowledge, occurred. \red{There is also an inherent tradeoff in throughput due to photonic transmission and component losses.} We note however that progress is ongoing in this area, particularly at the shorter end of the mid-infrared band (around 3-4~\textmu m) \citep[e.g.][]{2017Harry,2019Gretzinger}. In principle, all components could be photonic.

\section{Systematic instrumental uncertainties}

\label{sec:uncertainties}
In this section, we analyse the systematic uncertainties associated with this beam combiner implementation, and the effects of phase fluctuations of the input beams more broadly, on the robustness and sensitivity of the kernel-nulling architecture.

\subsection{Adaptive nullers and alignment procedure}
In this architecture, systematic errors can come from a number of places: errors in the phase and amplitude of the input beams, as well as errors in the optical elements of each beam splitter module. We can, however, eliminate some of these errors immediately through a careful calibration process using an adaptive nuller.

Adaptive nullers, as described by \cite{LayAdaptiveNull}, are compensators that can adjust the phase and amplitude of an input beam of light through the use of a deformable mirror (DM). The light is spectrally dispersed onto a DM, adjusted to tune the wavelength dependent phase and amplitude, before being dedispersed and recollimated for beam combination. Amplitude is tunable through phase tilts orthogonal to the dispersion direction, when combined with spatial filtering. This is an invaluable tool for correction of the beams on input, and has also been shown to provide stable achromatic phase shifts \citep{Peters2010}. 

For our purposes, the adaptive nuller can also be used to eliminate any errors in the top nulling row of beam splitter modules ($\vb{A}_1$ through $\vb{A}_4$) with a process similar to the following:
\begin{enumerate}
    \item Modulate the adaptive nuller on input $\vb{V}_1$ such that it maximises amplitude for all wavelengths. This input is set as the global phase reference.
    \item Modulate the adaptive nuller on input $\vb{V}_2$ so that, as well as entering the spatial filter and removing wavelength dependence, it forms a nulled output on output $\vb{W}_2$.
    \item Repeat stage two with the other three inputs, forming nulls on outputs $\vb{W}_3,\vb{W}_4$ and $\vb{W}_5$ respectively.
    \item If necessary, reduce the amplitude of input $\vb{V}_1$ and repeat steps 2 \& 3.
\end{enumerate}

In this manner, the phases of the input beams can be managed to completely remove any phase errors in the top nulling beam splitter optics, and any errors in the amplitudes of the inputs will just result in less overall light; the inputs can be modified such that they all have the amplitude of the dimmest input beam. \red{In other words, the beam splitter aberrations that may have an impact on the null and stellar leakage, including chromatic aberrations, can be removed by the adaptive nuller}. \red{This same process can also be used to intermittently remove alignment drifts between the nuller and fringe tracker during observations; this is discussed and proven to be sufficiently efficient in Appendix \ref{app:null_tuning}.} 

The spatial filters at the end of the first row are required to remove unwanted spatial modes of the light that would prevent deep nulls. \red{Any optical losses after the spatial filter will not have any impact on stellar leakage terms, and will affect zodiacal and exozodiacal light equally. These latter losses primarily affect the ability to form a good kernel output}.


\subsection{Beam combiner optical errors}
\label{sec:opticalerrors}

While the adaptive nullers are able to negate the effects of optical errors in the top nulling row of beam splitters, the remaining six modules will still contribute to errors in the kernel-nulls. To simulate this, we apply random fluctuations to the optical parameters based on a predefined root-mean-squared (RMS) error:
\begin{align}
    \theta &= \theta_0 + \frac{|\Delta R|}{\cos{\theta_0}}x\\
    \phi &= \phi_0 + \Delta\phi x,
\end{align}
where $x \in [-1,1]$ is is a random number, which is uniformly distributed to simulate the effect of a typical pass or fail optical specification. The parameters $\theta_0$ and $\phi_0$ are the true values as defined in Table \ref{tab:kernel-5_params}, $|\Delta R|$ denotes the error in the reflectance of the beam splitter, and $\Delta\phi$ represents the error in the phase shifter. For the remainder of this analysis, we consider three sets of these uncertainties and refer to the pair by their $|\Delta R|$ amount. These uncertainties are chosen for realistic manufacturing tolerances from optics suppliers:
\begin{align*}
    &&|\Delta R| &= 2\% & \Delta\phi &= 1\degree&\\
    &&|\Delta R| &= 5\% & \Delta\phi &= 3\degree&\\
    &&|\Delta R| &= 10\% & \Delta\phi &= 6\degree&.
\end{align*}

\begin{figure}
  \centering
  \begin{subfigure}{0.8\linewidth}
    \centering
    \includegraphics[width=\linewidth]{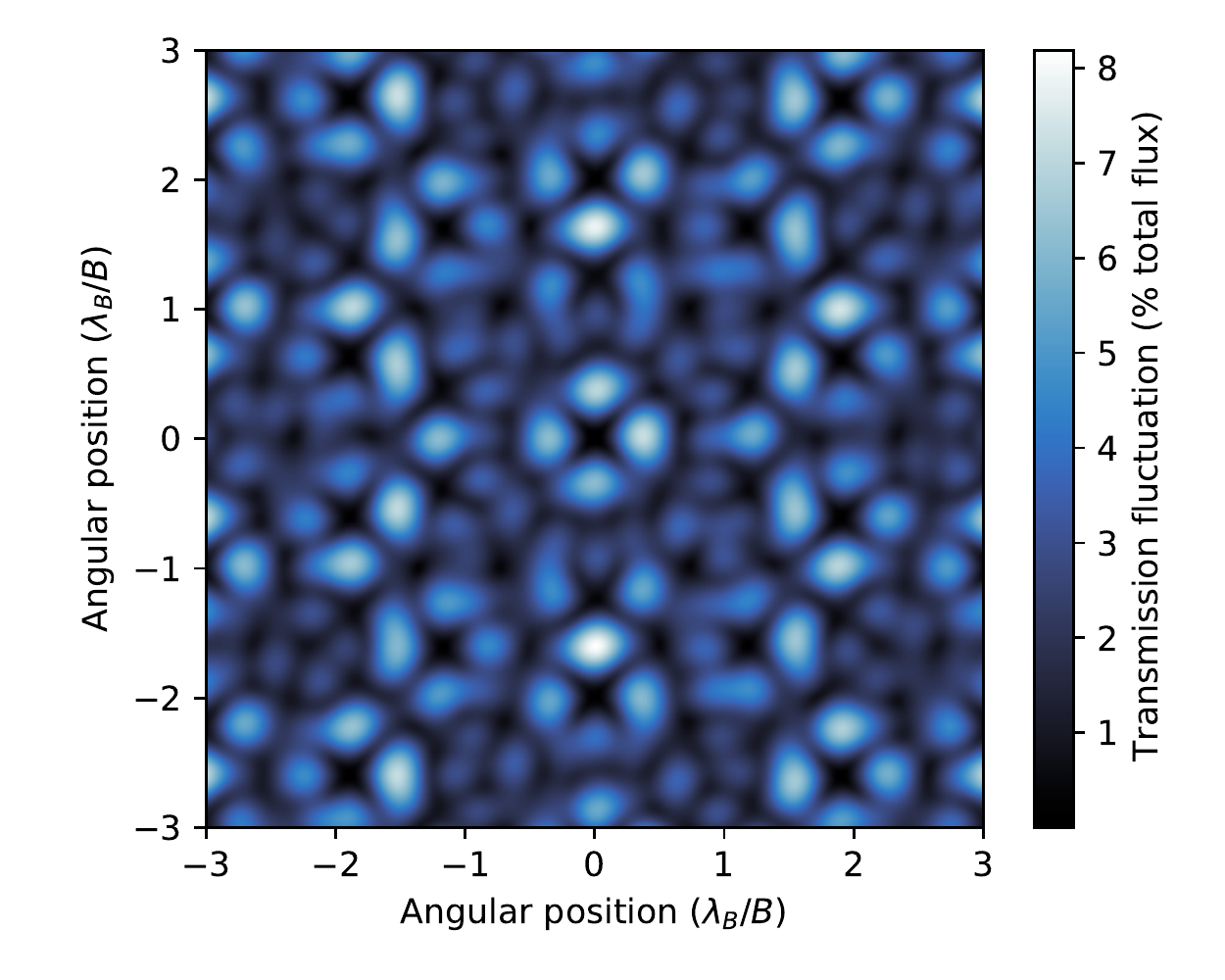}
   \caption{Kernel 1}
  \end{subfigure}
  \hfill
  \begin{subfigure}{0.8\linewidth}
    \centering
    \includegraphics[width=\linewidth]{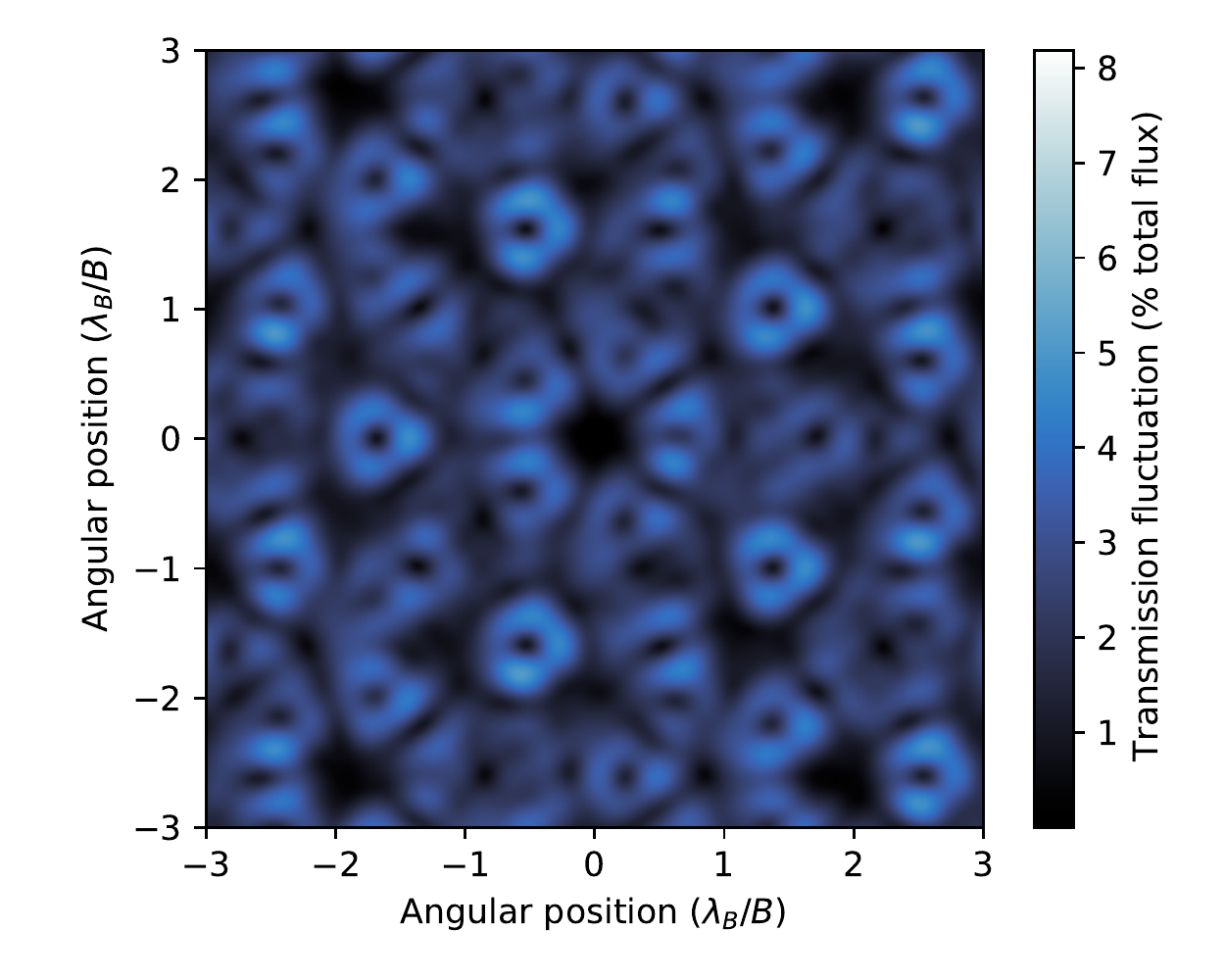}
   \caption{Kernel 2}
  \end{subfigure}
    \caption{Standard deviation of each kernel as a function of angular position, given as a percentage of the total array flux, for $|\Delta R|=5\%$.}
    \label{Img:trans_std}
\end{figure}

We ran a Monte-Carlo simulation to find the standard deviation of the kernel maps as a function of angular coordinate when these errors are applied. We assume a pentagonal arrangement of the telescopes equivalent to \red{the inset in Figure \ref{Img:kernel-5-guyon}}. The two maps, assuming $|\Delta R| = 5\%$, are shown in Figure \ref{Img:trans_std}. Here we see that the kernels show a maximum standard deviation of 8\% of the total telescope flux, with an average standard deviation of 2.7\% and 2.2\% of kernel 1 and 2 respectively.  We also find that the average standard deviations for $|\Delta R| = 2\%$ are 1\% and 0.9\%, and for $|\Delta R| = 10\%$ we have 5.7\% and 4.6\% respectively. 

\begin{figure}
    \centering
    \includegraphics[width=0.9\linewidth]{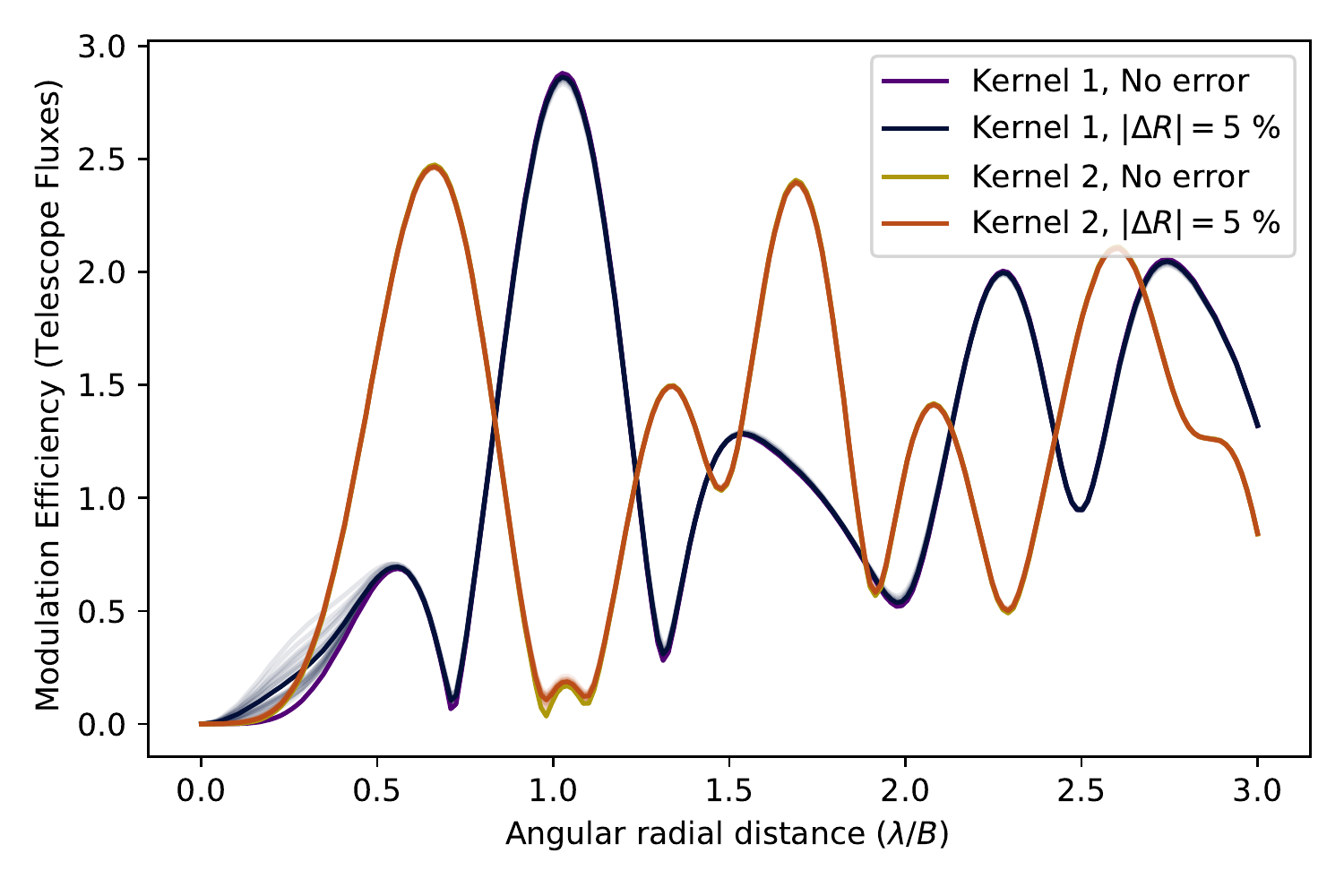}
    \caption{Modulation efficiency (RMS azimuthal average) of the kernel maps as a function of radius. The average with no beam combiner optical errors is overlaid on the average of twenty random draws with $|\Delta R|=5\%$. The random draws themselves are also plotted with low opacity.}
    \label{Img:radius_rms_std}
\end{figure}

We also examine the effect of these uncertainties on the modulation efficiency (RMS azimuthal average) of the kernel map; this highlights the fluctuations that may influence the power of the planet signal as the array rotates. We show this in Figure \ref{Img:radius_rms_std}, where we overlay the modulation efficiency as a function of radius with no error, normalised by flux per telescope, on top of the average of twenty random draws with $|\Delta R|=5\%$. What is apparent here is that with this amount of error, the modulation efficiency is not significantly affected, indicating the information in the signal does not significantly change with these optical errors even though the detailed map structure requires additional calibration or modelling. \red{For clarity, we note that Figure \ref{Img:radius_rms_std} plots the modulation efficiency of the kernel maps (that is, the difference between two outputs), and not the raw nulled outputs themselves; it is the latter that defines the order of the null and the amount of stellar leakage.}

\subsection{Null depth}
\label{sec:null_depth}
\begin{figure}
  \centering
  \begin{subfigure}{0.9\linewidth}
    \centering
    \includegraphics[width=\linewidth]{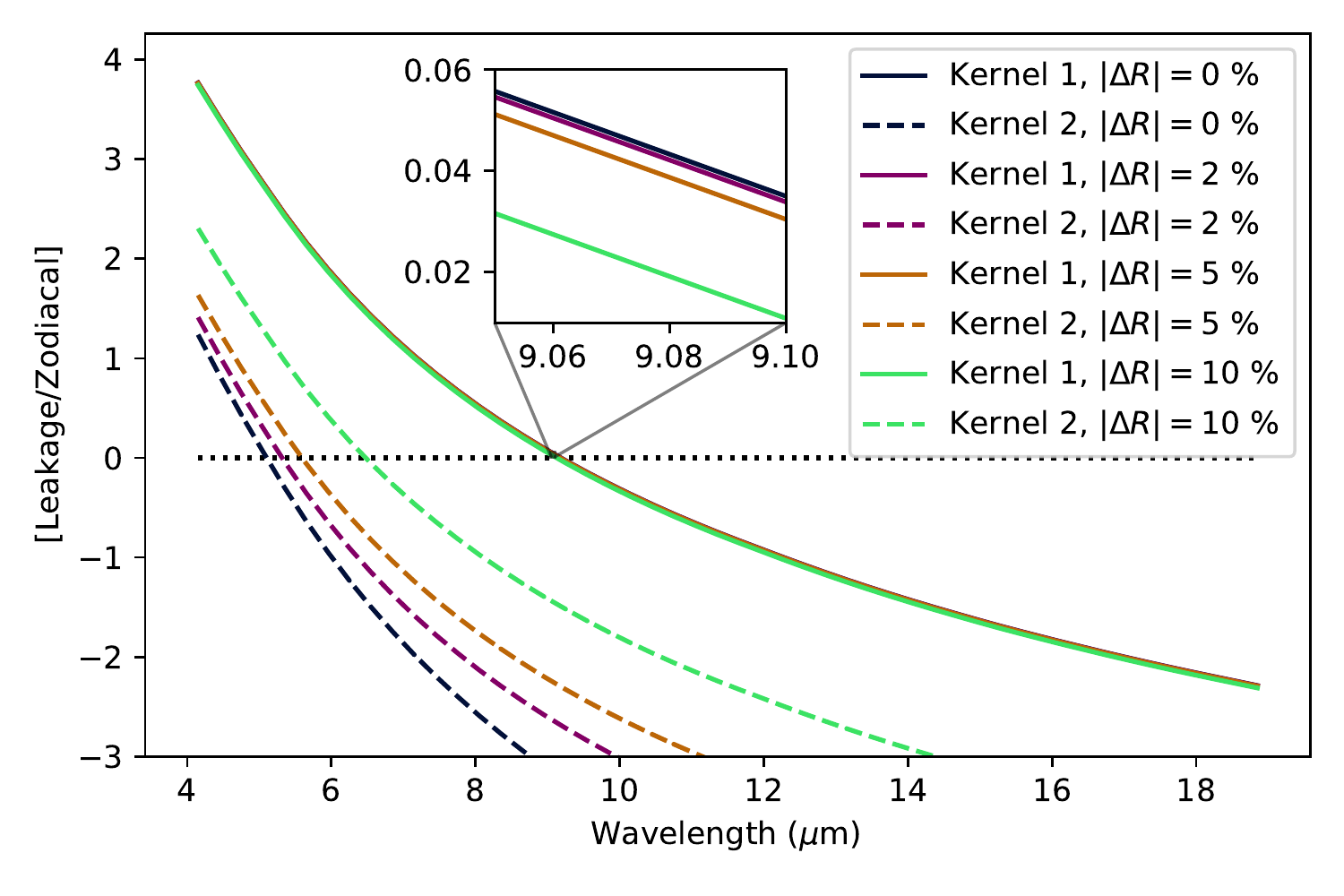}
    \caption{M5V type star at 5~pc}
  \end{subfigure}
  \hfill
  \begin{subfigure}{0.9\linewidth}
    \centering
    \includegraphics[width=\linewidth]{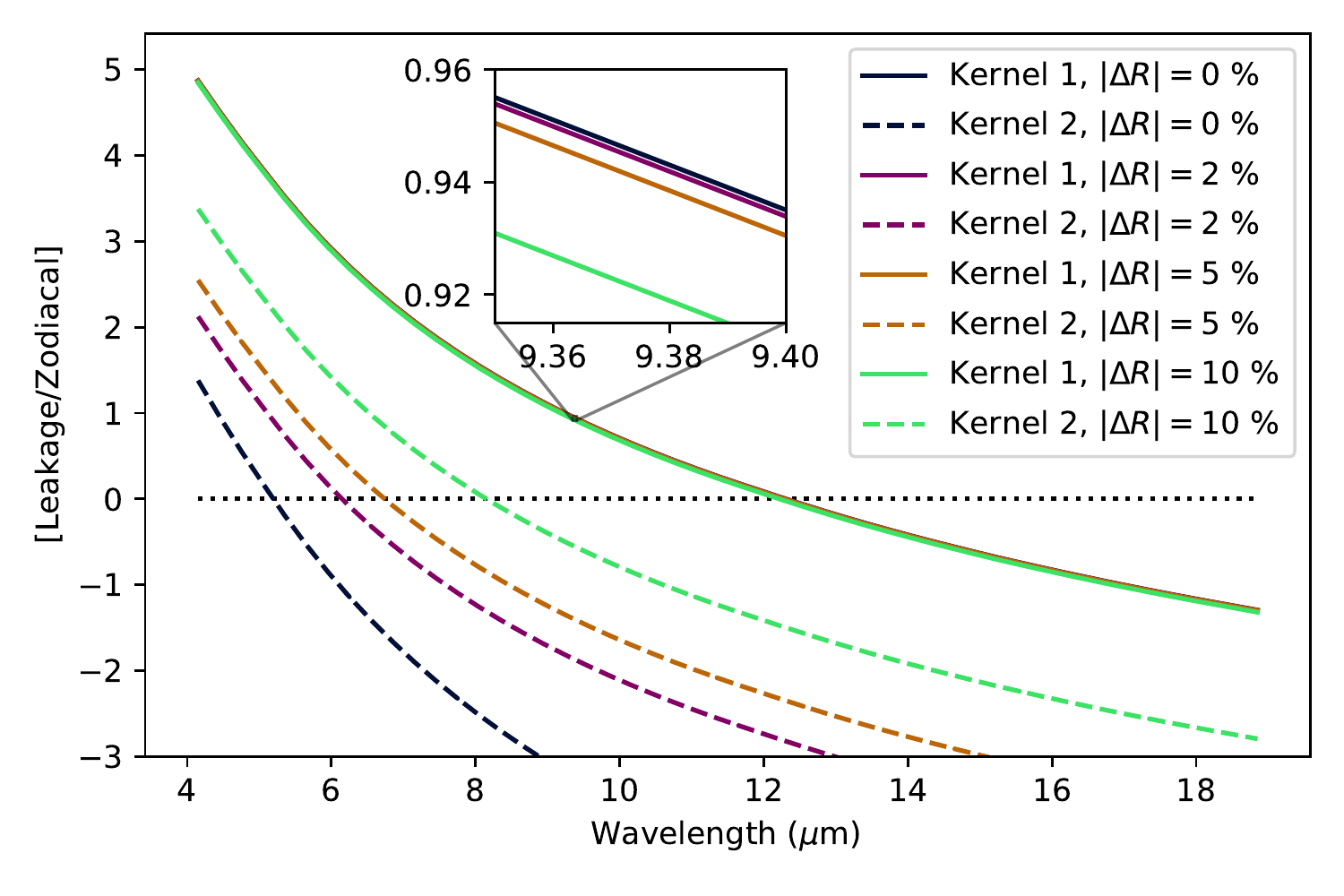}
    \caption{G2V type star at 5~pc}
  \end{subfigure}
    \caption{Base-10 logarithm of the ratio of the stellar leakage to zodiacal light against wavelength for two different stars and varying amounts of beam combiner optical error. The black dashed line divides the upper region where the combiner is dominated by stellar leakage, and the lower region where the instrument is zodiacal limited. \red{The inset highlights the proximity of the different optical error lines for kernel 1.}}
    \label{Img:leakage_zodiacal}
\end{figure}

From the plots in Figure \ref{Img:radius_rms_std}, the most concerning trend induced by errors in the optical elements is the effect on the null; whether the null no longer reaches the desired depths and thus greatly increasing stellar leakage. Using the simulation machinery detailed in LIFE4, we calculated the base-10 logarithm of the ratio of the stellar leakage noise and zodiacal background light as a function of wavelength, assuming a 2~m aperture size. The wavelength range chosen is between 4 and 19~\textmu m, to align with that of \cite{LIFE1} and LIFE4. We calculated these plots for two stars located at 5~pc: a M5V dwarf based on Proxima Centauri, and a G2V dwarf based on the Sun. The latter was chosen based on the closest stars of F or G stellar type: there are three stars within 6~pc ($\tau$ Ceti, e Eri and $\eta$ Cas) and can be considered on average to be roughly a solar-type star at 5~pc. Hence this can be used as an extreme scenario on stellar leakage. 
\red{For the local zodiacal light, we once again used the JWST background calculator\footnote{GitHub: \url{https://github.com/spacetelescope/jwst_background}}, assuming sky coordinates equivalent to Tau Boo (as an average stellar case). This is roughly equivalent to $1\times10^{-7}$ the blackbody radiation of a 300~K source.} The simulation was then repeated for $|\Delta R|$ values of 2\%, 5\% and 10\%, to see the effect of optical errors on stellar leakage. The plots are shown in Figure \ref{Img:leakage_zodiacal}.

We can see from these plots that kernel 2 is \red{far} more sensitive to these optical errors; as kernel 1 is only a second-order null, it is \red{much} more dominated by stellar leakage, and as such the noise floor is above that induced by these beam combiner optical errors. The fourth-order null of kernel 2, however, produces a smaller stellar leakage term and as such, a reduction in the null depth arising from optical errors. We find that an error of 2\% results in a six fold increase in stellar leakage for the solar type star, with a similar increase at 10\% for the M-dwarf. 

\red{One important point to be made here is that these optical errors will not modify the total amount of stellar leakage present in the system. This can be seen in the zoomed-in inset of Figure \ref{Img:leakage_zodiacal}, where additional optical abberations marginally decreases the amount of stellar leakage present in kernel 1, counteracting the increase in kernel 2. This is due to the optical errors only affecting the mixing component of the beam combiner, and not the nulling stage (the first row of beam splitting units). With 90\% optical error (that is, essentially random values for all the mixing beam splitter units), the two kernel curves overlap, converging slightly below the position of the kernel 1 line in Figure \ref{Img:leakage_zodiacal}. The main effect these errors induce then, is reducing the effectiveness of kernel 2 (which can produce a much higher S/N) and removing the robustness against systematic errors that kernel outputs are designed to have (see Section \ref{sec:robust}).}

Furthermore, while this error results in quite a shift, particularly for the solar type star, we remind the reader that these stars are extreme scenarios. Most stars in the LIFE catalogue \citep{LIFE1} are further than 5~pc away, and the amount by which stellar leakage dominates at the short wavelengths decreases with distance. There are also few non M-dwarfs within 5~pc, and a 2\% \red{$|\Delta R|$} error impacting an M-dwarf measurement does not result in a large change in the leakage to zodiacal ratio. Nevertheless, this does indicate that optical errors in the beam combiner are important to minimise, particularly in suppressing stellar leakage at the short wavelengths. Conversely, \red{post-nulling} optical beam combiner errors \red{at typical commercial specifications of $\sim$ 10\%} do not matter beyond approximately 8~\textmu m as measurements will be strongly zodiacal background dominated. 

\subsection{Null stability}
In LIFE4, we provided a simple approximation of the RMS fringe tracking requirements to remain limited by photon background noise, rather than fluctuations in the null. We found that the interferometer should aim for < 9~nm RMS when looking at M-dwarfs, and around 2~nm for G-dwarfs, both at about 5~pc. We now look further into this, including the impact of optical beam combiner errors on the fringe tracking requirements and null stability.
 
From equation 50 in LIFE4, recall that the minimum fringe tracking error needed to remain photon noise limited can be estimated from this equation:
\begin{equation}
    \langle\phi^2\rangle F_\text{star} < \max{\left[\frac{P_\text{zodiacal}}{A}, F_\text{leakage}\right]}.
\end{equation}
where $F_\text{star}$ is the stellar flux, $P_\text{zodiacal}$ is the zodiacal light per diffraction limited mode (in phot/s), $F_\text{leakage}$ is the stellar leakage flux and $A$ is the single telescope aperture.

\begin{figure}
  \centering
  \begin{subfigure}{0.9\linewidth}
    \centering
    \includegraphics[width=\linewidth]{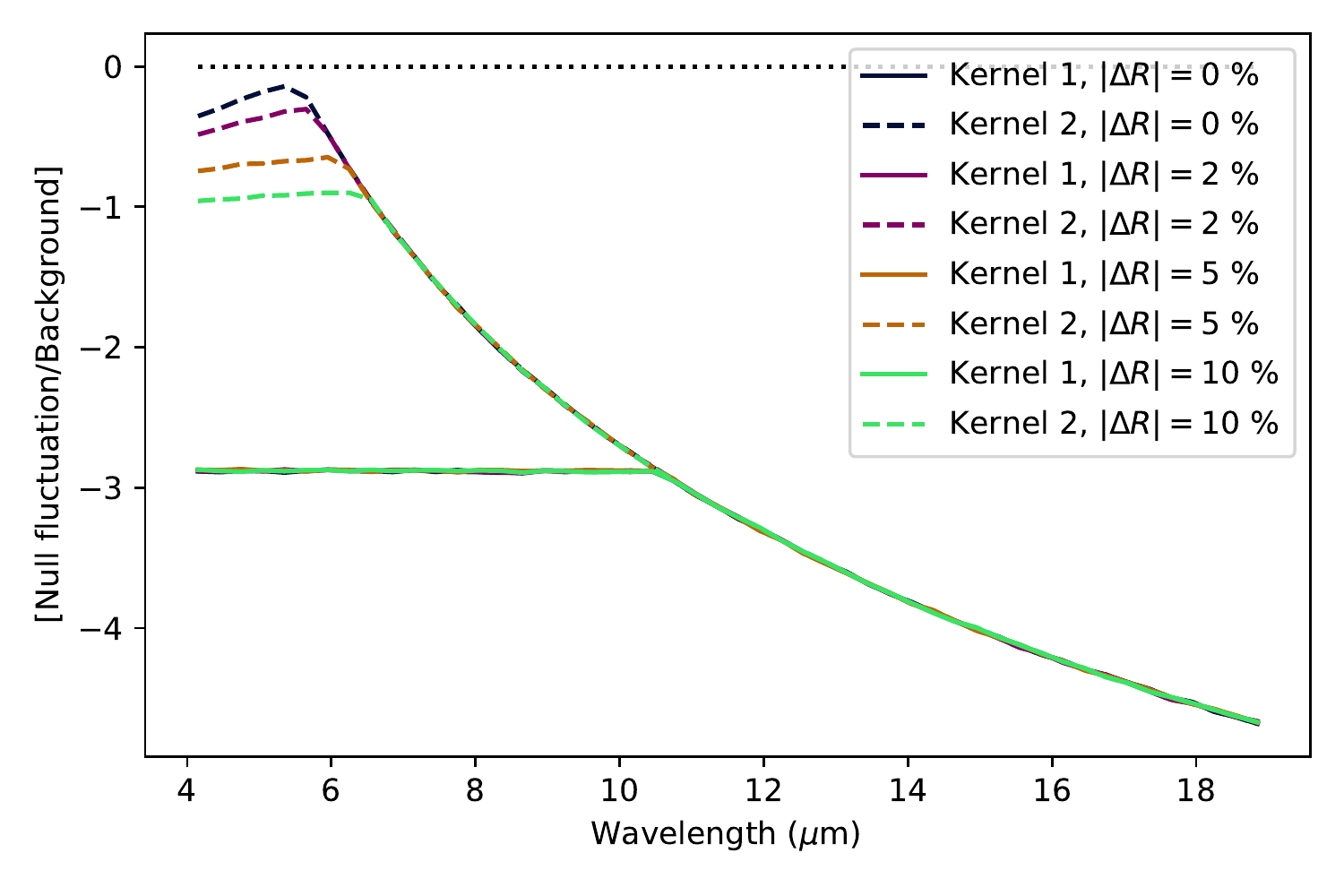}
   \caption{M5V type star at 5~pc}
  \end{subfigure}
  \hfill
  \begin{subfigure}{0.9\linewidth}
    \centering
    \includegraphics[width=\linewidth]{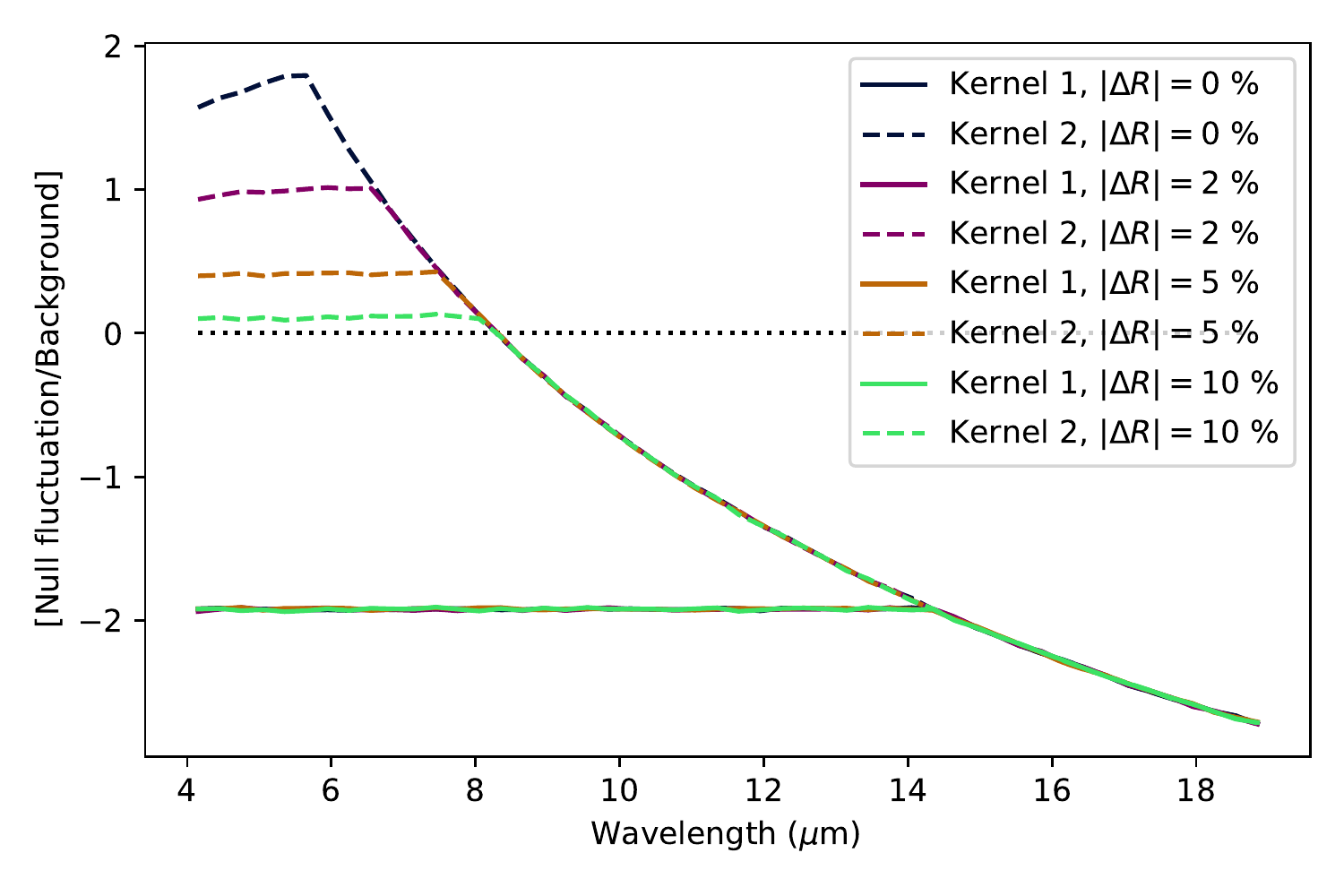}
   \caption{G2V type star at 5~pc}
  \end{subfigure}
    \caption{Base-10 logarithm of the ratio of the null fluctuation noise to background noise against wavelength for a fringe tracking RMS of $\delta = 5$~nm. The background noise is chosen to be the maximum of stellar leakage and zodiacal light for that given wavelength. Plotted for two different stars and varying amounts of beam combiner optical error. The black dashed line divides the upper region where the combiner is dominated by instrumental errors induced by fluctuations in the null depth, and the lower region where the instrument is photon limited by either stellar leakage or the zodiacal background.}
    \label{Img:null_fluct}
\end{figure}

We simulate a random erroneous phase of all five telescopes, adding a term
\begin{align}
\label{eq:random_phase}
    \phi_i = \frac{2\pi}{\lambda}X,
\end{align}
where $i$ represents the index of the telescope, $\lambda$ the wavelength in nanometers and $X\sim\mathcal{N}(0,\delta)$ is a random variable pulled from a normal distribution with zero mean and a standard deviation of $\delta$, the RMS fringe tracking error in nanometers. We then calculate the mean square response over a large number of random phases, $\langle\phi_i^2\rangle$ and multiply by the stellar flux to obtain the noise due to null fluctuations. In Figure \ref{Img:null_fluct}, we plot as a function of wavelength the base-10 logarithm of the ratio of the null fluctuations against the maximum background of that given wavelength. This was plotted for the same two stars and optical beam combiner errors as for Figure \ref{Img:leakage_zodiacal}. An RMS fringe tracking error of $\delta = 5$~nm was \red{used for} this plot.

Firstly, we note that as in Figure \ref{Img:leakage_zodiacal}, kernel 1 is never dominated by the null fluctuations, again due to it not providing as sensitive a null. We can also see the areas for which stellar leakage and zodiacal light dominate as a function of wavelength: dominant stellar leakage is represented by a flat line (as leakage follows the same functional form as the stellar flux), and the downwards curve represents being zodiacal light dominated. As to be expected from the previous discussion, we see that with a greater beam combiner optical error, the stellar leakage dominates for a larger part of the spectrum. The stronger leakage also washes out the effect of null fluctuations: with a high optical error, these terms dominate and as such null stability becomes less important. 

\begin{table}[]
\centering
\caption{Minimum fringe tracking requirements to remain photon limited over 4 to 19 \textmu m for different stellar types and different optical error amounts at 5~pc.}
\label{tab:null_fluct}
\begin{tabular}{@{}cccc@{}}
\toprule
\multicolumn{1}{c}{\multirow{2}{*}{\begin{tabular}[c]{@{}c@{}}Beam combiner\\ optical error $|\Delta R|$\end{tabular}}} & \multicolumn{3}{c}{Fringe tracking RMS requirement ($\delta$)} \\
\multicolumn{1}{c}{} & G2V star & K2V star & M5V star \\ \midrule
0\% & 0.7~nm & 1~nm & 6~nm \\
2\% & 1.5~nm & 2~nm & 7.5~nm \\
5\% & 3~nm & 4~nm & 11~nm \\
10\% & 4.5~nm & 5.5~nm & 15~nm \\ \bottomrule
\end{tabular}%
\end{table}

In Table \ref{tab:null_fluct}, we plot the minimum fringe tracking RMS needed to remain dominated by photon noise at all wavelengths for stars of type G, K and M at 5~pc, as well as for differing amounts of beam combiner optical error. Interestingly, the RMS fringe tracking requirement is stricter than the previous calculation, being less than 1~nm when looking at G-dwarfs with no optical errors and about 6~nm for an M-dwarf. A balance needs to be struck with regards to acceptable stellar leakage and the achievable fringe tracking uncertainties. With $|\Delta R|$ = 5\%, the additional stellar leakage is not increased too much, and the fringe tracking requirement remains manageable at 3~nm; allowing measurements for G, K and M type stars to remain photon noise limited at all wavelengths. 

We also note here that the instrument remains zodiacal limited regardless of beam combiner optical errors for $\lambda > 8$~\textmu m and $\delta = 5$~nm even for solar-type stars at 5~pc. The teams defining the science goals for the LIFE mission will need to take this into consideration: performance will be unequal throughout the spectral range, with spectral signatures beyond 8~\textmu m requiring less stringent RMS fringe tracking uncertainties due to being zodiacal background dominated in this regime. 

\subsection{Sensitivity and robustness of the kernel}
\label{sec:robust}
We also looked at the effect of these optical errors on the robustness of the kernel itself. As described in \cite{2018Martinache}, the kernel operators are designed to be independent of systematic errors in phase, and are analogous to the technique of `phase chopping' in the literature \citep[e.g.][]{Woolf97,Lay04,Mennesson05}  with regards to removing unwanted symmetrical emission such as exozodiacal light and the removal of instrumental errors. However, with the inclusion of beam combiner optical errors the outputs are no longer `pure' kernels and will hence be affected by systematic phase offsets. To analyse this, we once again modified the phase of the telescopes as in Equation \ref{eq:random_phase} and took the standard deviation of the resulting kernel output from this phase error. We interpret the error introduced in the phase as a systematic piston offset; that is, over multiple exposures, the phase at the centre of the map will average to a non-zero value. \red{We note that these phase errors are separate and fundamentally different to the internal optical phase shift errors in the previous sections, primarily as these will affect the stellar leakage and nulling stage of the combiner.}

\begin{figure}
  \centering  \begin{subfigure}{0.9\linewidth}
    \centering
    \includegraphics[width=\linewidth]{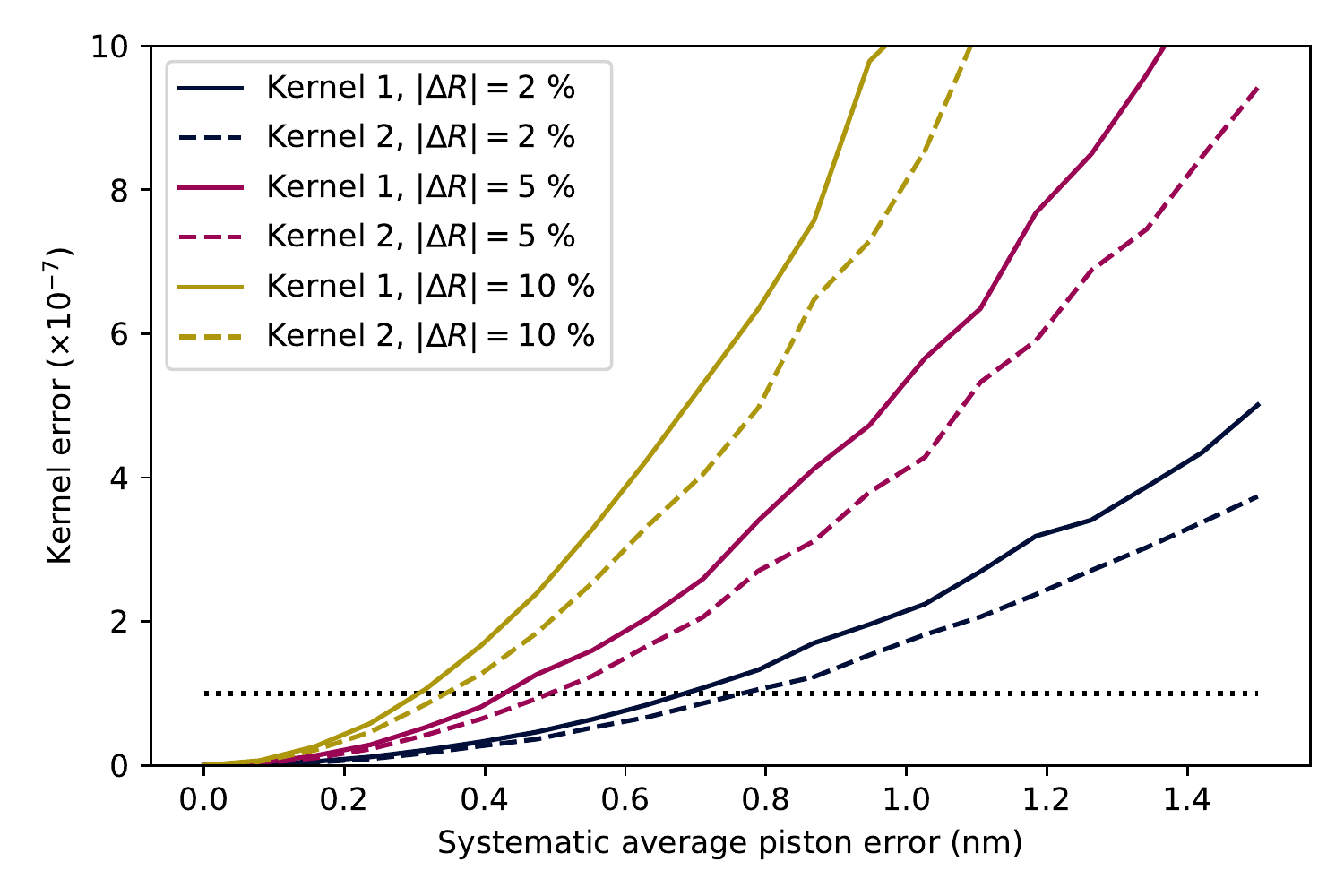}
    \caption{$\lambda$ = 4~\textmu m}
  \end{subfigure}
  \hfill
  \begin{subfigure}{0.9\linewidth}
    \centering
    \includegraphics[width=\linewidth]{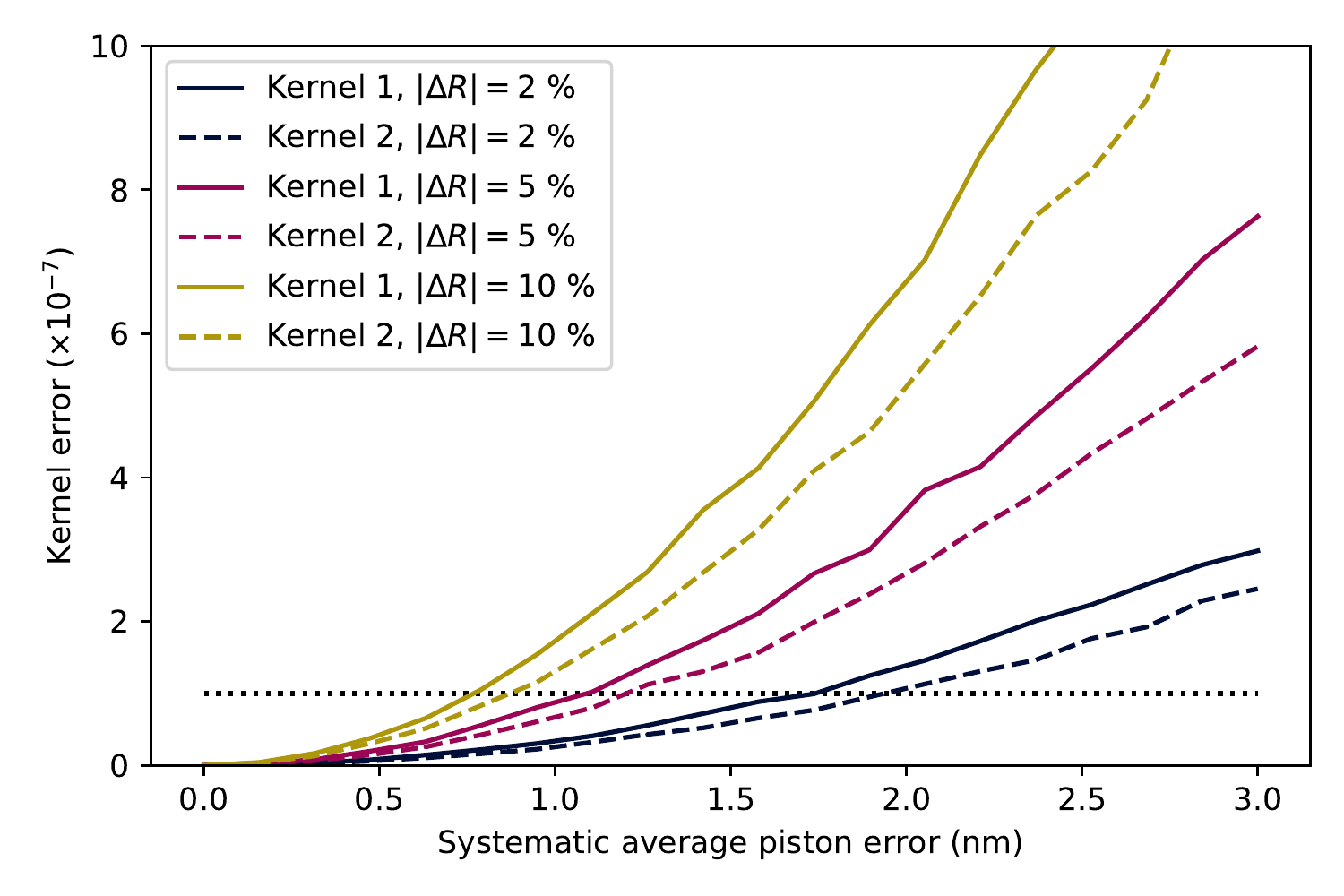}
    \caption{$\lambda$ = 10~\textmu m}
    \label{Img:phase_err_10um}
  \end{subfigure}
  \hfill
    \caption{Error in the kernel, plotted against systematic piston error in nanometers for a variety of optical beam combiner errors. The dotted line at $1\times 10^{-7}$ represents the point where the kernel should be sensitive enough to detect an Earth-like planet around a solar type star.}
    \label{Img:phase_err_plot}
\end{figure}

We plot this systematic piston offset (in nanometers) against the standard deviation of the kernel outputs for multiple beam combiner errors in Figure \ref{Img:phase_err_plot}. We also plotted the curve for two different wavelengths: one at 10~\textmu m, and one at 4~\textmu m. We can identify that the curves follow a quadratic relationship with the systematic piston error, and that kernel 1 is marginally more sensitive to systematic piston error than kernel 2. 

In order for a confident planet detection to be made, we need to make sure that the kernel output is sensitive enough to the planet and that systematic errors will not show up as false signals. In the mid-infrared, the contrast for an Earth-like planet against a solar-type star is $1\times 10^{-7}$ \citep{Defrere2018}, and so we need to ensure the error in the kernel lies below this amount. This is shown on the plots as a black dotted line. From this, it is apparent that systematic piston error needs to be kept as low as possible, especially at short wavelengths and with large optical errors. At $\lambda = 4$~\textmu m, an optical error of $|\Delta R| = 2\%$ requires a systematic piston error < 0.75~nm, whereas $|\Delta R| = 10\%$ requires a very stringent 0.3~nm or less. The longer wavelength plot \red{at $\lambda = 10$~\textmu m} is more lenient, suggesting a systematic error of < 1.8~nm at $|\Delta R| = 2\%$ and 0.7~nm for $|\Delta R| = 10\%$. This finding emphasises a result that has been consistent across all of these investigations: the shorter wavelengths are much more affected by instrumental errors.

\begin{figure}
    \centering
    \includegraphics[width=0.9\linewidth]{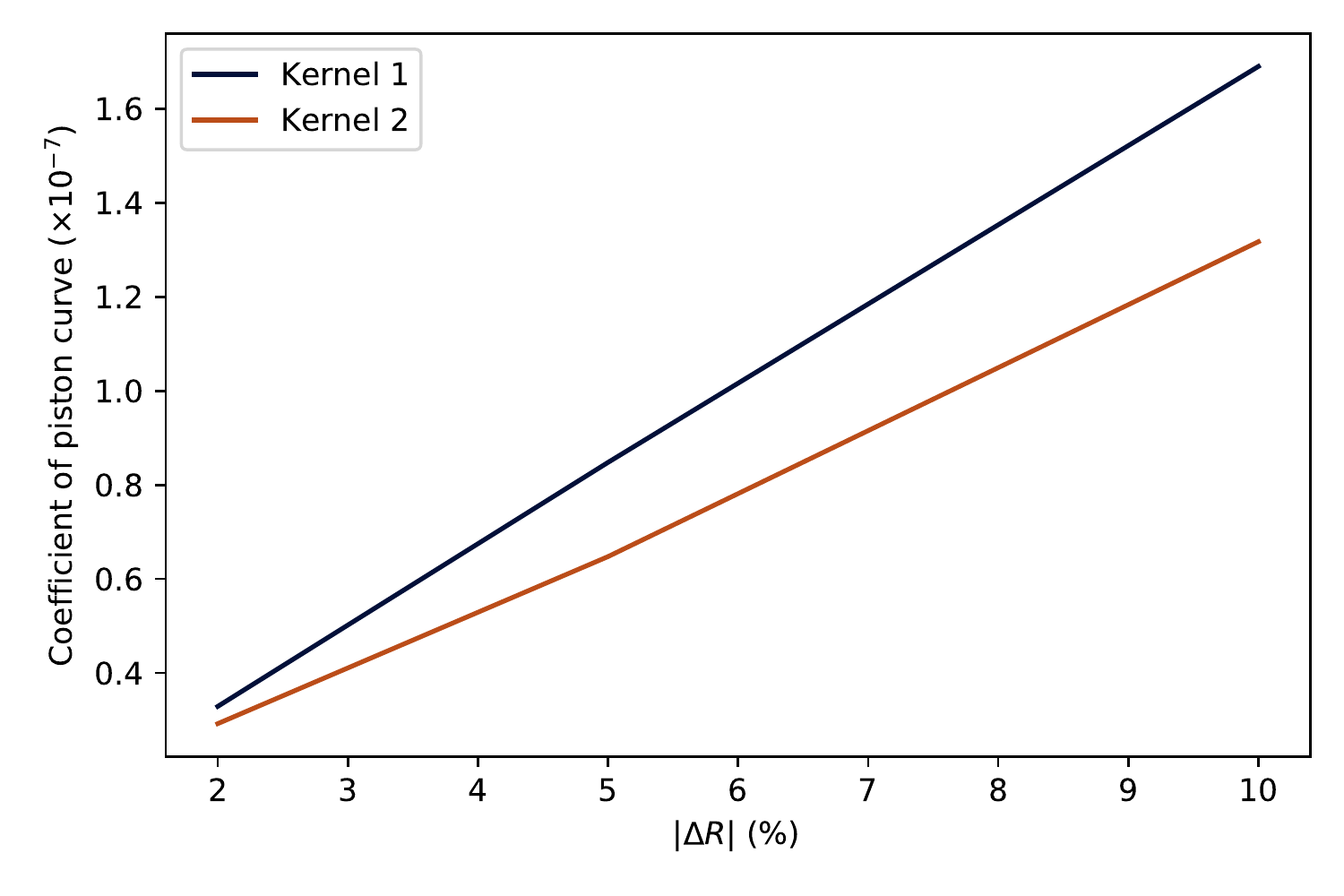}
    \caption{Relationship between the quadratic coefficients of Figure \ref{Img:phase_err_10um} and their associated error in the beam combiner $|\Delta R|$.}
    \label{Img:phase_err_coeff_plot}
\end{figure}

We also fitted a quadratic to each of the curves in Figure \ref{Img:phase_err_10um}, and plotted the coefficient against the respective error amount $|\Delta R|$. This is shown in Figure \ref{Img:phase_err_coeff_plot}. We identify that the coefficient of the quadratic curve scales linearly with beam combiner error, and hence show that the kernel error is overall third order in systematic piston errors. 

\subsection{Phase chopping}
\label{sec:phase_chop}

While kernel-nulling allows us to remove on-axis symmetric photon noise sources, as well as being resistant to second-order errors in piston \citep{2018Martinache}, there are still residual instrumental noise sources that could be removed through phase chopping, namely \red{differential zodiacal background levels in each output, and} detector noise. This technique involves rapidly swapping the rows \red{corresponding to kernel output pairs with each other (i.e., for the five-telescope combiner, swapping rows two and three, and four and five)}. In doing so, the kernel output remains the same but the signals are being measured on different detector pixels. Hence we can remove any slowly variable detector bias or gain effects in the system. 

\red{Furthermore, if there is a different amount of zodiacal light in each output due to different telescope sensitivities or similar effects, this will remain stationary in the output despite phase chopping and can hence be removed. It is this latter point that is the main motivation for phase chopping: for an Earth-like planet around a solar analogue at 5~pc, the zodiacal light can dominate the signal by three orders of magnitude at the upper end of the bandpass. Hence differing coupling of starlight and zodiacal background at the 1\% level can cause major difficulties in signal extraction and so should be removed if possible.}

Phase chopping with our beam combiner design is theoretically not difficult; all that is required is to flip the signs for each of the phase shifts in front of beam splitter modules $\vb{A}_5$ through $\vb{A}_{10}$. Mathematically:
\begin{align}
    \phi_i \Longrightarrow -\phi_i && i\in [5,10].
\end{align}
This could be made to happen, for example, by putting \red{beam splitters} on piezo stages and rapidly moving them by a fraction of a wavelength to induce a rapid phase shift sign flip (and hence a `\red{delay} chop'). 

Of course, by chopping in delay, this induces another error term. As we are working over a large wavelength range and these changing phase shifts will only work at specific wavelength, elsewhere in the bandpass will incur a degree of chromatic phase error. This can be minimised by reducing the size of the bandpass (such as the use of multiple beam trains as described in Section \ref{sec:implementation}), by making the reference wavelength for phase chopping in the centre of the bandpass \red{(such that the errors on either end of bandpass are equal)}, and by designing the beamsplitter to have a wavelength independent 0 or $\pi$ phase shift. This last point allows us to halve the amount of phase shift error that would otherwise occur, as well as ensure that both outputs on either side of the phase chop have symmetrical errors. 

To model this error, we split our nominal wavelength range into three \red{sub-bandpasses}, evenly spaced with regard to the amount of phase error induced at the end of the sub-bandpasses. These ranges became 4-6.7~\textmu m, 6.7-11.2~\textmu m and 11.2-19~\textmu m. \red{The shift in delay ($\delta$) required to perform the phase chop is given by
\begin{align}
    \delta_i = \Delta \phi_i\frac{\lambda_c}{2\pi} = \text{min}(2\phi_i,2\pi - 2\phi_i)\times\frac{\lambda_c}{2\pi} && i\in[5,10],
 \end{align}
where $\lambda_c$ is the central (reference) wavelength of the relevant sub-bandpass and $\Delta \phi_i$ is the minimum change in phase required to flip the sign of the $i$th beam splitter phase shift. As previously mentioned, we defined the central wavelength to be such that the chromatic errors at the edges of the sub-bandpasses are equal. For the ranges listed above, these wavelengths are $\lambda_c =$ 5.02~\textmu m, 8.43~\textmu m and 14.17~\textmu m. The chromatic error in phase at wavelengths other than the central wavelength can then be calculated by
\begin{align}
    \sigma_{\phi_i} = \frac{1}{2}\left|\Delta \phi_i - \frac{2\pi\delta_i}{\lambda}\right|
    = \frac{1}{2}\left|\Delta \phi_i\left(1 - \frac{\lambda_c}{\lambda}\right)\right|.
\end{align}}

\begin{figure}
  \centering
  \begin{subfigure}{0.9\linewidth}
    \centering
    \includegraphics[width=\linewidth]{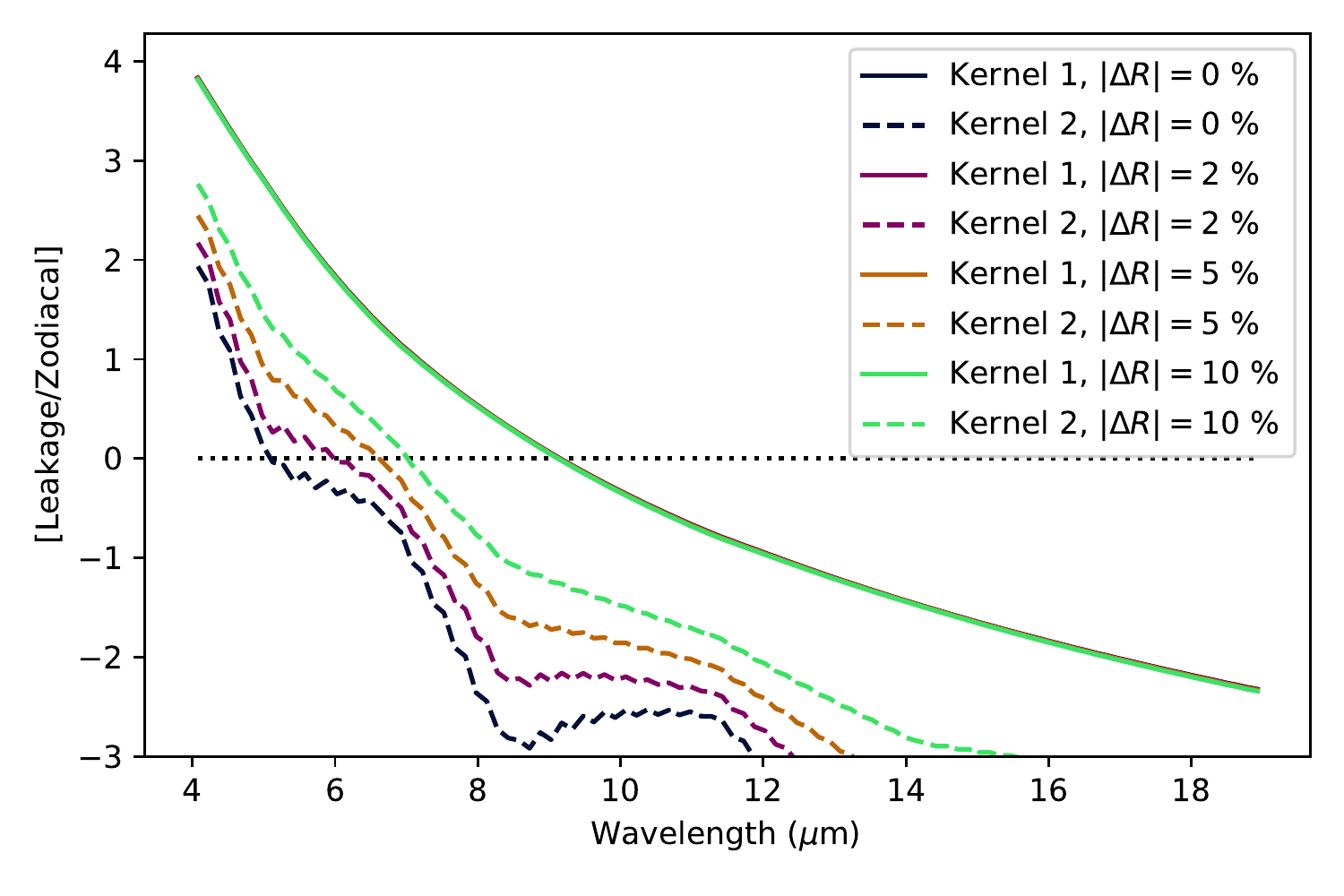}
    \caption{M5V type star at 5~pc}
  \end{subfigure}
  \hfill
  \begin{subfigure}{0.9\linewidth}
    \centering
    \includegraphics[width=\linewidth]{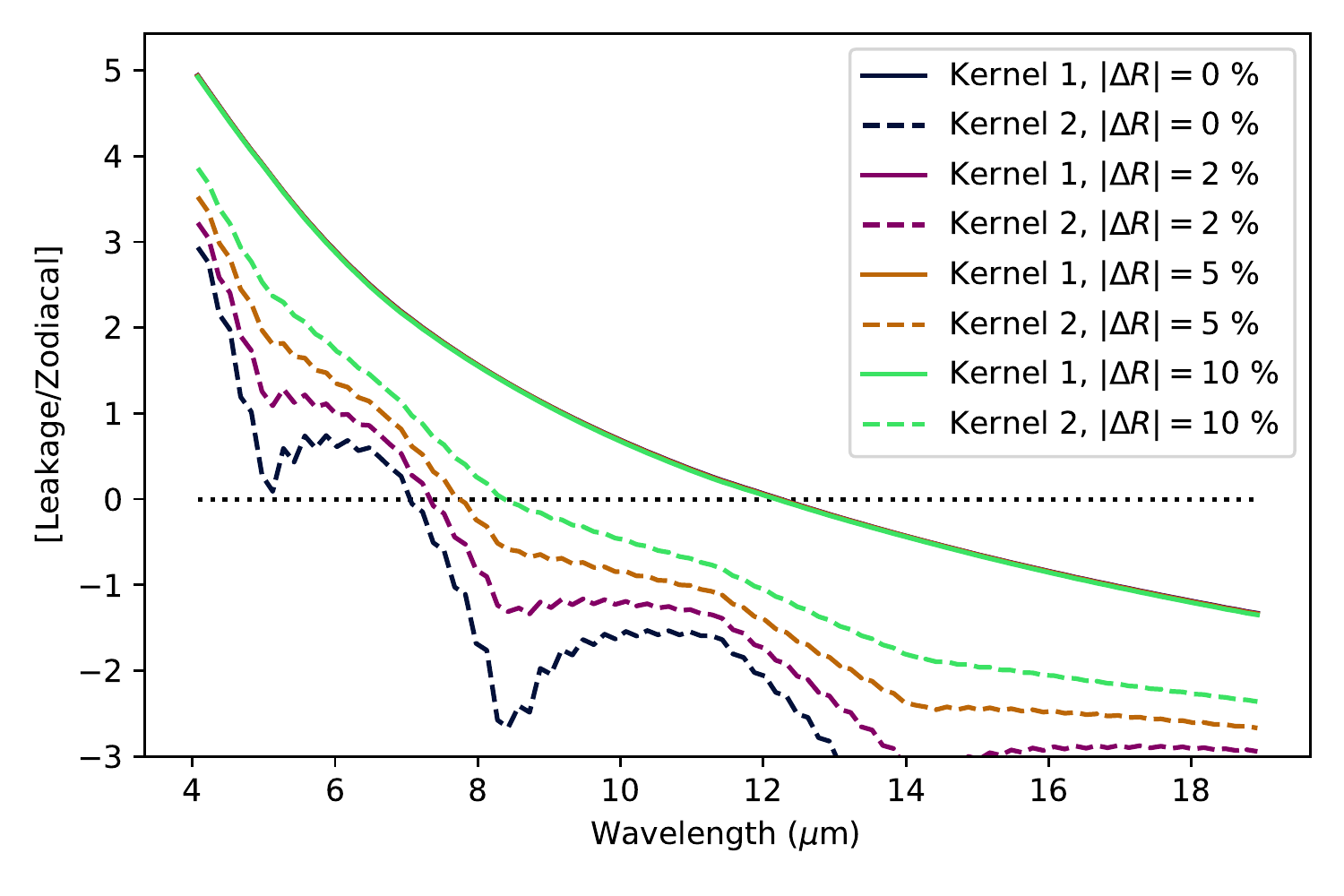}
    \caption{G2V type star at 5~pc}
  \end{subfigure}
    \caption{Base-10 logarithm of the ratio of the stellar leakage to zodiacal light against wavelength for two different stars, varying amounts of beam combiner optical error. Chromatic phase error has been induced by a delay chop.}
    \label{Img:leakage_zodiacal_phasechop}
\end{figure}

For the Kernel-5 beam combiner, these phase shifts and the maximum error associated with them at the edge of the sub-bandpass (in radians) are
\begin{align*}
    \Delta \phi_5 &= 2.17 &&& \sigma_{\phi_5} &= 0.28\\
    \Delta \phi_6 &= 0.60 &&& \sigma_{\phi_6} &= 0.08\\
    \Delta \phi_7 &= 1.62 &&& \sigma_{\phi_7} &= 0.21\\
    \Delta \phi_8 &= 0.29 &&& \sigma_{\phi_8} &= 0.04\\
    \Delta \phi_9 &= 0.99 &&& \sigma_{\phi_9} &= 0.13\\
    \Delta \phi_{10} &= \pi &&& \sigma_{\phi_{10}} &= 0.4
\end{align*}
To see the effect that this chromatic phase chop error would have on the measurements, we performed the same simulation as in Section \ref{sec:null_depth}, except adding the relevant phase chop error to the  $\Delta \phi$ of each beam splitter. This is shown in Figure \ref{Img:leakage_zodiacal_phasechop}.

We can see in these plots that the added chromatic error indeed makes the stellar leakage considerably worse for kernel 2, with kernel 1 being \red{barely} effected for the same reasons as in Section \ref{sec:null_depth}. We also see the effects of chromaticity - the leakage is at a minimum in the centre of each sub-bandpass (where there should be no added error) and increases to a local maximum or inflection point at the edges. The effect is quite strong at the shortest wavelengths, reducing the effect of the null by an order of magnitude. However, the second kernel is still zodiacal dominated beyond 8~\textmu m; hence this will only be a problem for the shortest wavelengths around the closest stars. 

If this were deemed to be too great an error to propagate uncorrected, we could add a thin wedge of glass to a second piezo stage in front of each of the effected beam splitters that could act as a corrector for this chromatic effect, \red{though} this doubles the number of piezos and would considerably increase the beam combiner's complexity. \red{Nevertheless, due to the likelihood of varying zodiacal background levels in each output, as well as detector effects, this added complexity is likely a worthwhile tradeoff.}  

\section{Redundancy for failed telescopes}
\label{sec:redundancy}
\subsection{Kernel-5 nuller}

One significant advantage of the `Guyon'-type beam combiner design for the Kernel-5 nuller described in Section \ref{sec:implementation}, on top of the planet yield advantages discussed in LIFE4, is the ability for it to continue producing robust observable measurements even if a collecting spacecraft fails. In other words, the Kernel-5 nuller will still be able to function with only four telescopes. This is not applicable to the traditional X-array beam combiner - if one of the telescopes of that design fails, the main mission objectives for detecting Earth-like exoplanets is severely compromised.

This safeguard against a damaged telescope can be implemented through the use of a well placed shutter in the midst of the beam splitters, shown as $\vb{S}$ in Figure \ref{Img:kernel-5-guyon}. If a collector telescope fails, all that is required is for the four operating telescopes to move into input positions two through five (that is, the failed telescope corresponds to input $\vb{V}_1$), and the shutter to close. We can emulate this in matrix notation through blocking beam one at the start of the relay (representing the failed telescope; $\vb{F}$) and then blocking beam two in between beam splitting modules $\vb{A}_4$ and $\vb{A}_7$ (representing the shutter; $\vb{S}_1$). Inserting these into equation \ref{eq:solve}:
\begin{align}
    \vb{F} = \begin{bmatrix}
    0 & 0 & 0 & 0 & 0\\
    0 & 1 & 0 & 0 & 0\\
    0 & 0 & 1 & 0 & 0\\
    0 & 0 & 0 & 1 & 0\\
    0 & 0 & 0 & 0 & 1
    \end{bmatrix} &&     \vb{S} = \begin{bmatrix}
    1 & 0 & 0 & 0 & 0\\
    0 & 0 & 0 & 0 & 0\\
    0 & 0 & 1 & 0 & 0\\
    0 & 0 & 0 & 1 & 0\\
    0 & 0 & 0 & 0 & 1
    \end{bmatrix}
\end{align}
\begin{equation}
    \label{eq:shutter}
    \vb{\tilde{M}} = \vb{B}\vb{A}_{10}\vb{A}_9\vb{A}_8\vb{A}_7\vb{A}_6\vb{A}_5\vb{S}\vb{A}_4\vb{A}_3\vb{A}_2\vb{A}_1\vb{F}.
\end{equation}

\begin{figure}
    \centering
    \includegraphics[width=0.8\linewidth]{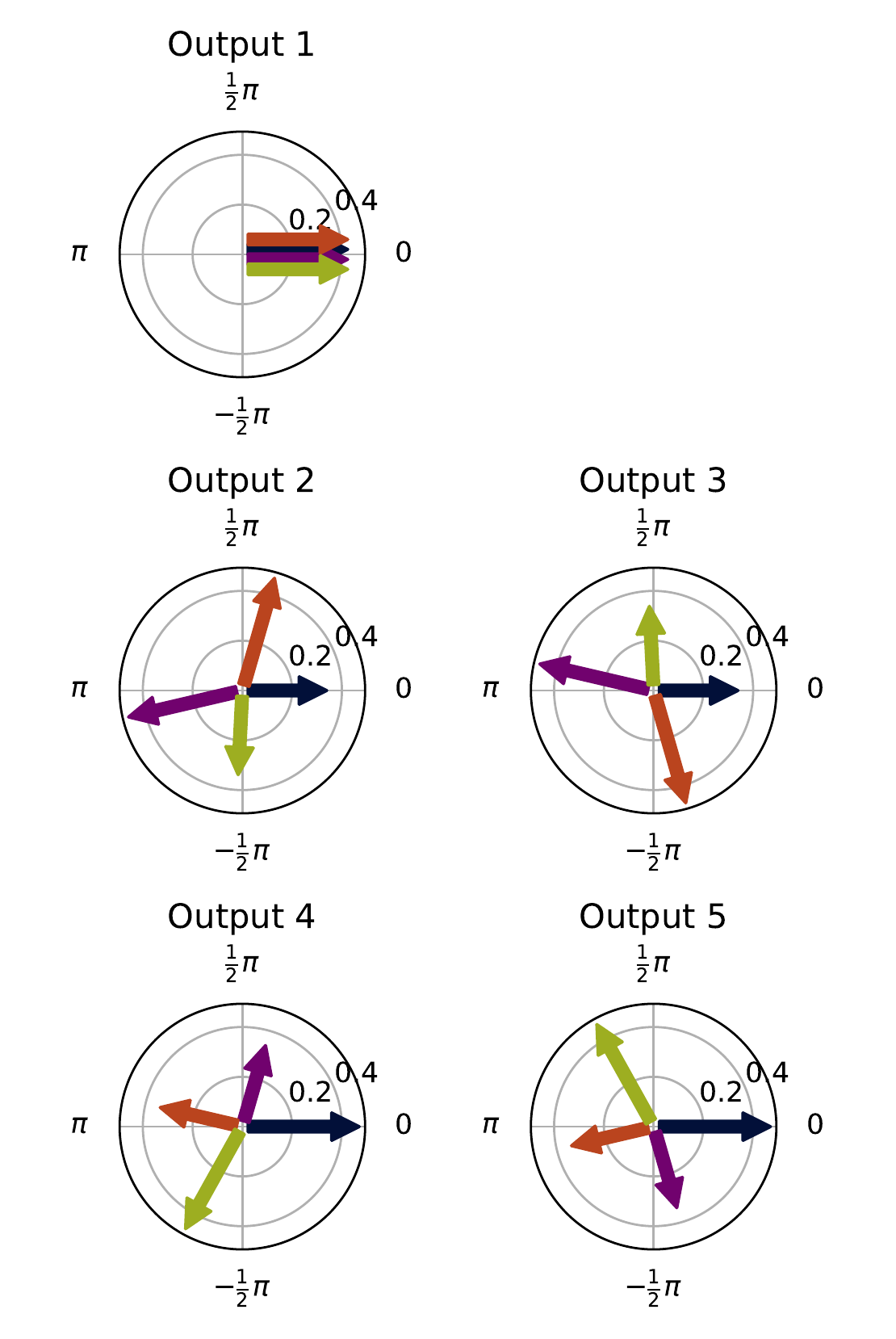}
    \caption{Complex Matrix Plot of the `damaged' Kernel-5 beam combiner with four telescope inputs.}
    \label{Img:CMP_K5}
\end{figure}

Using the same parameters of the beam splitters and phase shifters derived in section \ref{sec:implementation}, we can calculate the new transfer matrix ($\vb{\tilde{M}}$) for the damaged system:
\begin{equation}
       \vb{\tilde{M}} = \frac{1}{\sqrt{5}}\begin{bmatrix}
    0 & 1 & 1 & 1 & 1\\
    0 & -\frac{\sqrt{5}}{4} - ai & \frac{\sqrt{5}}{4} + bi & \frac{\sqrt{5}}{4} - bi & -\frac{\sqrt{5}}{4} + ai\\
    0 & -\frac{\sqrt{5}}{4} + ai & \frac{\sqrt{5}}{4} - bi & \frac{\sqrt{5}}{4} + bi & -\frac{\sqrt{5}}{4} - ai\\
    0 & \frac{\sqrt{5}}{4} + bi & -\frac{\sqrt{5}}{4} + ai & -\frac{\sqrt{5}}{4} - ai & \frac{\sqrt{5}}{4} - bi\\
    0 & \frac{\sqrt{5}}{4} - bi & -\frac{\sqrt{5}}{4} - ai & -\frac{\sqrt{5}}{4} + ai & \frac{\sqrt{5}}{4} + bi\\
\end{bmatrix},
\end{equation}
where
\begin{align*}
    a &= \sqrt{\frac{5}{2(5+\sqrt{5})}} &&&
    b &= \frac{1}{2}\sqrt{\frac{1}{2}(5+\sqrt{5})}.
\end{align*}
It is apparent from this system that output $\vb{W}_1$ is again the bright output. What is less apparent is that outputs $\vb{W}_2$ and $\vb{W}_3$, and $\vb{W}_4$ and $\vb{W}_5$ form enantiomorphic pairs; an attribute that allows them to form a kernel-null \citep{Laugier2020}. To demonstrate this, we perform a relative phase shift at the output (that is, change $\omega$) so that the contribution of input 2 is always real; this should result in the kernel-null pairs becoming complex conjugates of each other. We plot the pairs of outputs in a `Complex Matrix Plot', akin to \cite{Laugier2020}, in Figure \ref{Img:CMP_K5}, where it is easily seen that the pairs are mirror images of each other. Thus, even with one telescope no longer working, the system is able to produce two kernel-null outputs. 

\begin{figure}
  \centering
  \begin{subfigure}{0.75\linewidth}
    \centering
    \includegraphics[width=\linewidth]{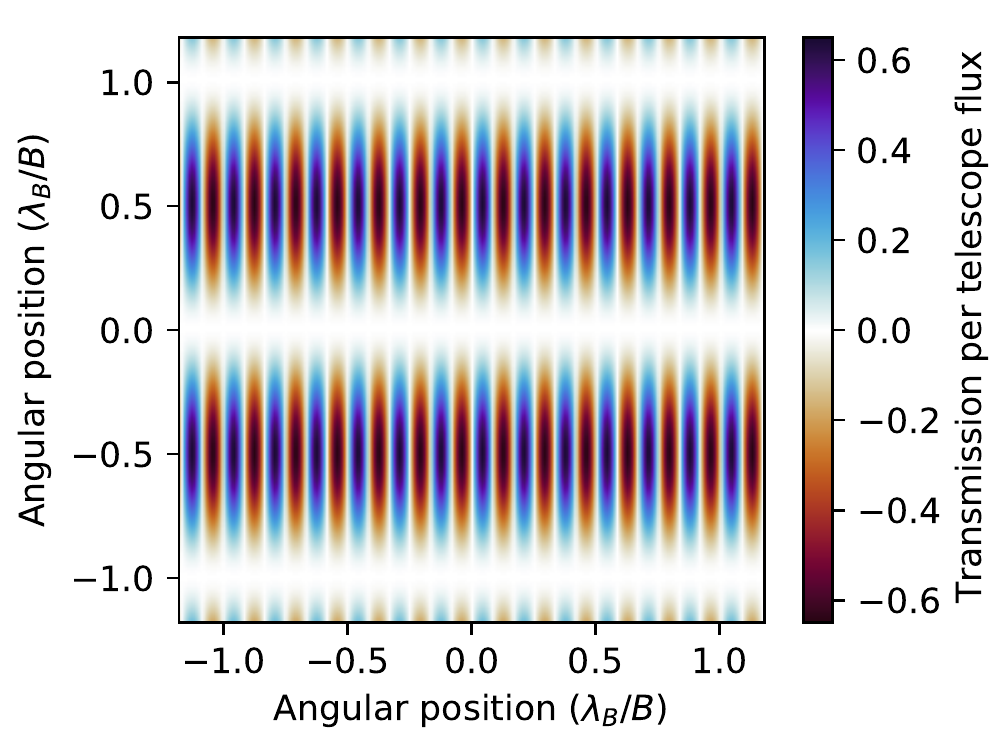}
    \caption{Kernel 1}
  \end{subfigure}
  \hfill
  \begin{subfigure}{0.75\linewidth}
    \centering
    \includegraphics[width=\linewidth]{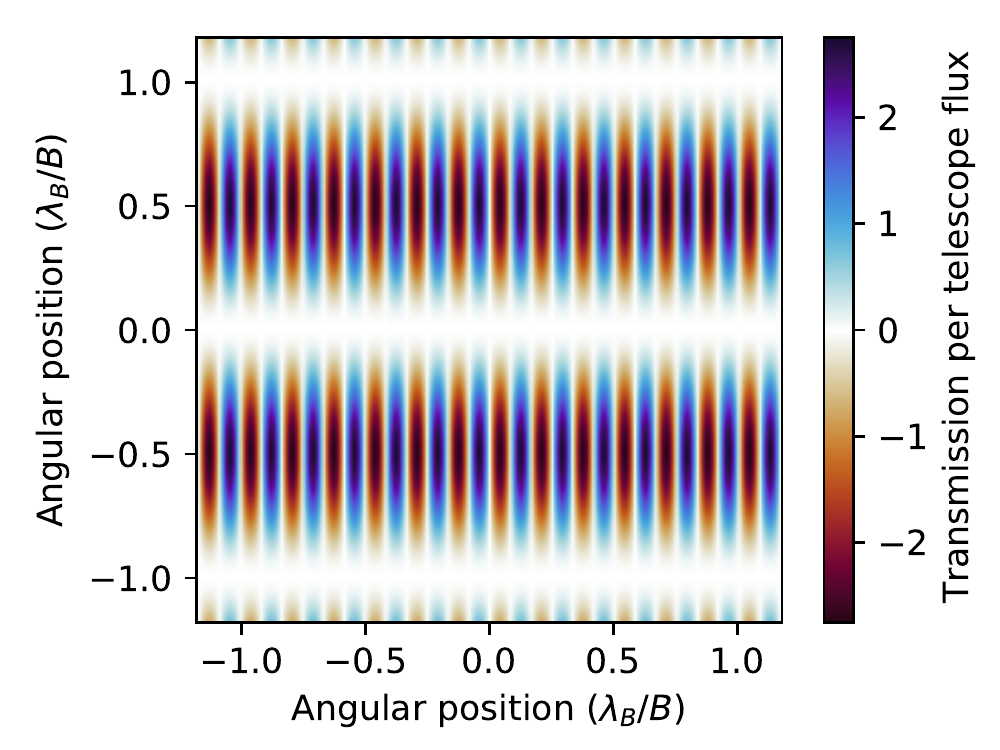}
    \caption{Kernel 2}
  \end{subfigure}
    \caption{Kernel maps of the `damaged' Kernel-5 beam combiner with four telescope inputs. It is important to note the different scaling in the colour maps; this is due to a combination of the beam combiner architecture, along with the geometry of the array.}
    \label{Img:kernel54Map}
\end{figure}

We show the output kernel maps in Figure \ref{Img:kernel54Map}, where we have assumed that the remaining four telescopes have changed configuration into a 6:1 X-array formation as in LIFE4 and \cite{Lay2006}. We find that one kernel produces a maximum transmission of 0.65 single telescope fluxes, and the other producing 2.75 telescope fluxes (together producing an efficiency of 85\% compared to the X-array, or 68\% with respect to the original 5 telescopes). 

\red{To properly compare the various combiners, let us consider a \red{flux-normalised} S/N metric defined as the ratio of the final background-subtracted S/N to that of a single telescope observing the faint planet in a background-limited imaging mode. For an array with $m$ telescopes, the upper bound for this value of this metric is $m$. If the output is in chopped pairs, then the upper bound is $m/\sqrt{2}$, and we can write for a single chopped pair:
\begin{align}
S/N_{i} &= \frac{m_\text{signal}}{\sqrt{2 f_\text{back}}},
\end{align}
where $m_\text{signal}$ is the number of telescopes that direct planet flux out a single output (equivalently the maximum value of the transmission map), and $f_\text{back}$ is the fraction of a single telescope background directed out a single output. This has an upper limit of 1, as the output is a single spatial mode. $f_\text{back}$ can be reduced by cold shutters.

For multiple output pairs that have uncorrelated noise, we can make a inverse variance weighted average of the planet flux signals, with a total signal to noise being
\begin{align}
S/N_\text{total} &= \sqrt{\sum_i \left(S/N_{i}\right)^2}.
\end{align}
\red{This is derived in Appendix \ref{app:SNR_metric}.} So, for the `shuttered' beam combiner, noting that the background here is reduced to $f_\text{back} = 0.8$ due to the shutter, we find that the total S/N metric is
\begin{equation}
    S/N = \sqrt{\left(\frac{2.75}{\sqrt{2\times0.8}}\right)^2 + \left(\frac{0.65}{\sqrt{2\times0.8}}\right)^2} = 2.23.
\end{equation}
The relative S/N of the damaged array is therefore 63\% of the original five telescope combiner (3.54), or 80\% of the equivalent non-damaged four-telescope array architecture (2.83)}. This infers that this configuration, made out of necessity due to a collector telescope failure, results in a \red{37\% S/N reduction compared to the original beam combination} architecture with 100\% of its telescopes functioning. \red{Despite this loss in S/N, this is still much better than the 100\% reduction that would occur in the X-array architecture due to it not being able to null if a telescope malfunctioned}. We note that the reason kernel 2 specifically contains most of the transmission is determined solely by the arrangement and numbering of telescopes in the array; a different arrangement would result in kernel 1 having the maximum transmission. 

\begin{figure}
    \centering
    \includegraphics[width=0.8\linewidth]{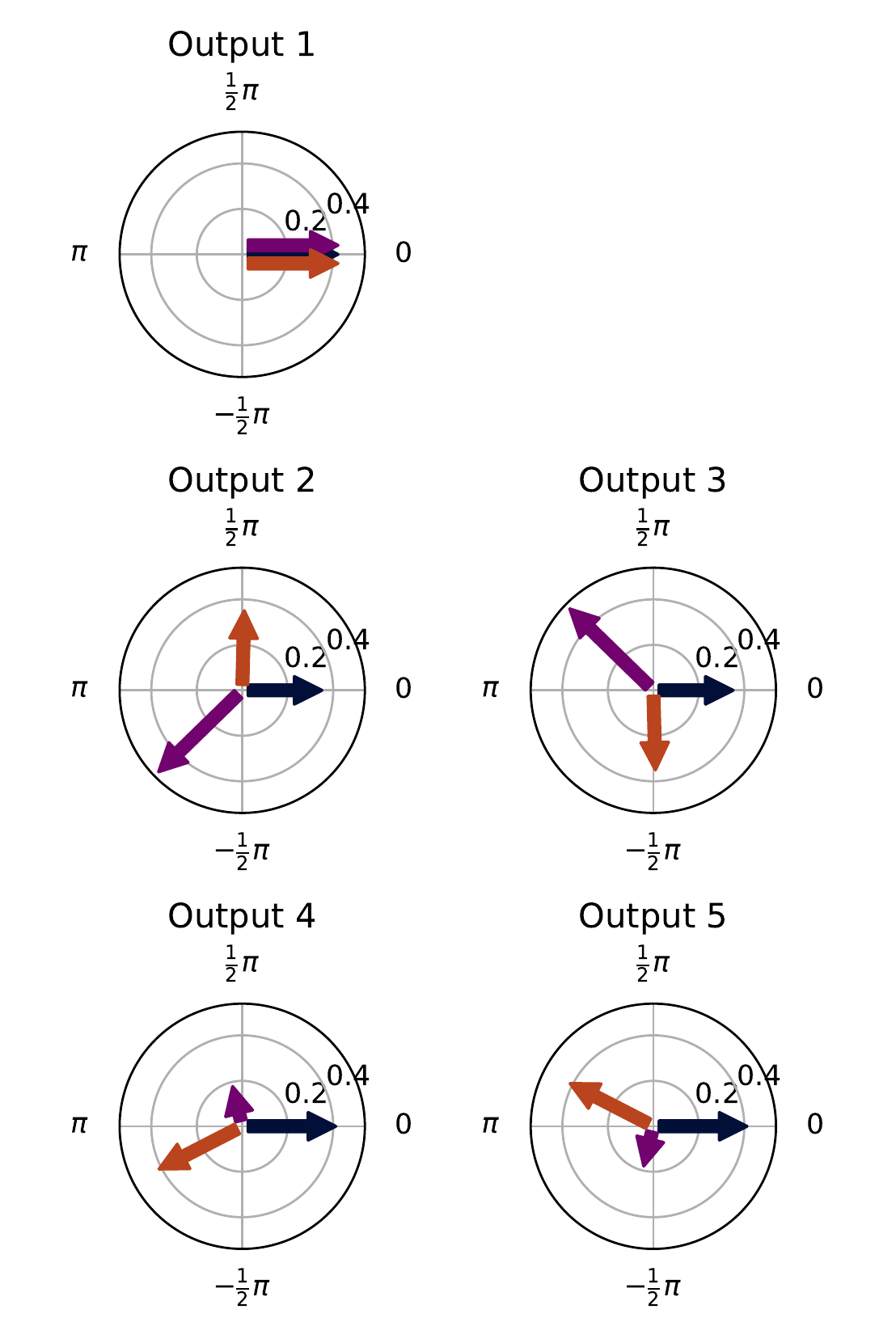}
    \caption{Complex Matrix Plot of the `damaged' Kernel-5 beam combiner with three telescope inputs.}
    \label{Img:CMP_K5_3}
\end{figure}

This idea can be extended further to three telescopes (that is, two telescopes failing) by implementing a second shutter ($\vb{S}_2$) between $\vb{A}_3$ and $\vb{A}_6$. This results in a transfer matrix of
\begin{equation}
       \vb{\tilde{M}} = \frac{1}{\sqrt{5}}\begin{bmatrix}
    0 & 0 & 1 & 1 & 1\\
    0 & 0 & \frac{\sqrt{5}}{6} + ai & \frac{\sqrt{5}}{6} - ci & -\frac{\sqrt{5}}{3} + ei\\
    0 & 0 & \frac{\sqrt{5}}{6} - ai  & \frac{\sqrt{5}}{6} + ci & -\frac{\sqrt{5}}{3} - ei\\
    0 & 0 & -\frac{\sqrt{5}}{6} + bi  & -\frac{\sqrt{5}}{6} - di & \frac{\sqrt{5}}{3} - fi\\
    0 & 0 & -\frac{\sqrt{5}}{6} - bi  & -\frac{\sqrt{5}}{6} + di & \frac{\sqrt{5}}{3} + fi\\
\end{bmatrix},
\end{equation}
where numerically
\begin{align*}
    a &= 0.7551280988643292 &&& d &= 0.2707664135274219\\
    b &= 0.9048040910575242 &&& e &= 0.3918568348616487\\
    c &= 1.1469849337259779 &&& f &= 0.6340376775301025.
\end{align*}
Again, we find that there are two sets of enantiomorphic pairs, shown in a CMP in Figure \ref{Img:CMP_K5_3}. The resultant maps have a maximum transmission of 0.91 and 1.47 telescope fluxes, and hence an array efficiency of 79\% compared to the Kernel-3 design, or 48\% compared to the original Kernel-5 design. The shutters also effectively reduce the background of the outputs by a factor of 0.6, which results in an effective S/N 1.58. \red{This is 44\% of the S/N of the original five-telescope combiner (3.54), or} 74\% of the S/N of a three-telescope combiner with all telescopes functioning (2.12). This modified combiner would therefore still be adequate to continue the mission after a failure of two spacecraft.

\subsection{Modified X-array}
\label{sec:modXarray}
While we stated that the X-array design does not allow for this redundancy advantage, this is only the case for the traditional beam combiner design consisting of two combiners with a $\pi$ phase shift along the nulled baseline, and then a $\frac{\pi}{2}$ phase chop of these nulled outputs. The X-array or Bracewell design could in fact be implemented in the same `Guyon'-type beam combiner as described in Section \ref{sec:implementation}. 

\begin{figure}
    \centering
    \includegraphics[width=0.9\linewidth]{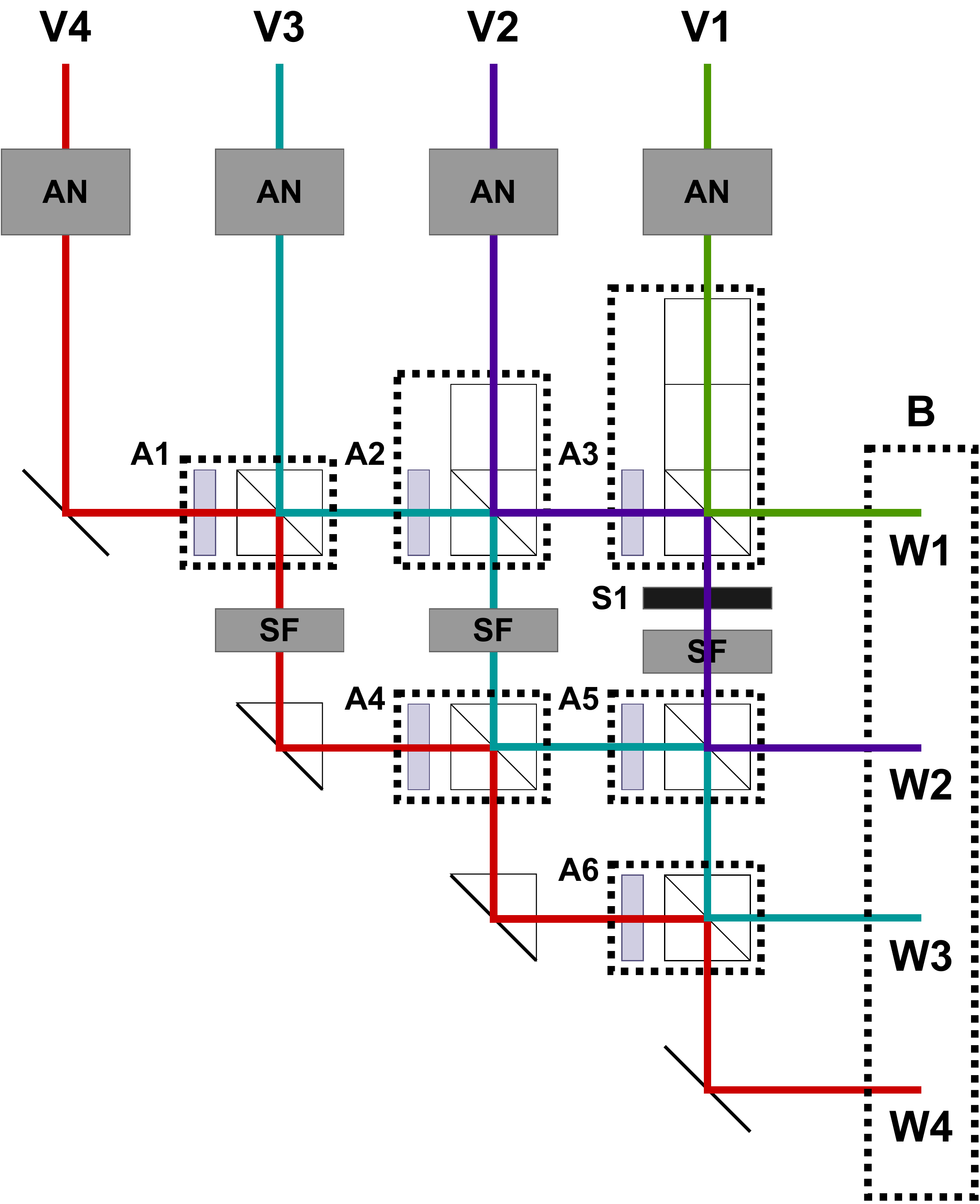}
    \caption{Schematic of an X-array beam combiner based on the design of \cite{Guyon2013}. The design is the same as in Figure \ref{Img:kernel-5-guyon}, except with four inputs and outputs, and the optical parameters found in Table \ref{tab:x-array_params}.}
    \label{Img:x-array-guyon}
\end{figure}

Consider a combiner with $m=4$ inputs and $n = 6$ beam splitter modules, shown in Figure \ref{Img:x-array-guyon}, along with the phase shifts and reflectance parameters found in Table \ref{tab:x-array_params}. Other than the parameters and number of inputs and outputs, this design in identical to that of the Kernel-5 nuller described in Section \ref{sec:implementation}. When the parameters are inserted into equation \ref{eq:solve}, we obtain the following transfer matrix:
\begin{equation}
       \vb{M} = \frac{1}{\sqrt{4}}\begin{bmatrix}
    1 & 1 & 1 & 1 \\
    1 & i & -i & -1\\
    i & 1 & -1 & -i\\
    -1 & 1 & 1 & -1
\end{bmatrix}.
\end{equation}

\begin{table*}[]
\centering
\caption{Optical parameters for the beam combiner design of the X-array, discussed in Section \ref{sec:modXarray} and displayed in Figure \ref{Img:x-array-guyon}}
\label{tab:x-array_params}
\begin{tabular}{@{}cccc@{}}
\toprule
              & Mixing Angle ($\theta$)                            & Reflectance coefficient $|R|$              & Phase Shift ($\phi$)                                  \\ \midrule
$\vb{C}_1$    & -$\frac{\pi}{4}\approx-0.785$                                   & $\frac{1}{\sqrt{2}}\approx0.707$                     & $\pi$                                                 \\
$\vb{C}_2$    & $\arcsin{\left(\frac{1}{\sqrt{3}}\right)}\approx0.615$                      & $\frac{1}{\sqrt{3}}\approx0.577$                     & $\pi$                                                 \\
$\vb{C}_3$    & $\frac{5\pi}{6}\approx2.618$                                   & $\frac{1}{2}$                            & $\pi$                                                 \\
$\vb{C}_4$    & $\pi-\arcsin{\left(\sqrt{\frac{5}{8}}\right)}\approx2.230$          & $\sqrt{\frac{5}{8}}\approx0.791$  & $\arctan{(2)}\approx1.107$ \\
$\vb{C}_5$    & $-\arcsin{\left(\frac{1}{\sqrt{3}}\right)}\approx-0.615$ & $\frac{1}{\sqrt{3}}\approx0.577$ & $\arctan{(3)}\approx1.249$              \\
$\vb{C}_6$    & $-\frac{\pi}{4}\approx-0.785$    & $\frac{1}{\sqrt{2}}\approx0.707$  & $\frac{3\pi}{4}\approx2.356$\\ \bottomrule
\end{tabular}%
\end{table*}

The middle two nulled rows of this matrix is equivalent to the middle two rows of the transfer matrix of the traditional X-array beam combiner found in equation 6 of LIFE4, with \red{an alternative} numbering of the telescopes. Hence this beam combiner could be used to produce the properties of the X-array as described in previous works \citep{LIFE1, LIFE4}. The benefit of this combination scheme is two-fold: there is an additional nulled output (albeit not a contribution to the kernel-null), and the same redundancy benefits as the Kernel-5 nuller apply. If a shutter ($\vb{S}_1$) is placed between modules $\vb{A}_3$ and $\vb{A}_5$, and defining the new transfer matrix in the same way as equation \ref{eq:shutter}, we obtain the following `damaged' transfer matrix:
\begin{equation}
       \vb{\tilde{M}} = \frac{1}{\sqrt{4}}\begin{bmatrix}
    0 & 1 & 1 & 1\\
    0 & \frac{1}{3} + i & \frac{1}{3} - i & -\frac{2}{3}\\
    0 & \frac{1}{3} -i & \frac{1}{3} + i & -\frac{2}{3}\\
    0 & \frac{2}{3} & \frac{2}{3} & -\frac{4}{3}
\end{bmatrix}.
\end{equation}

\begin{figure}
    \centering
    \includegraphics[width=0.8\linewidth]{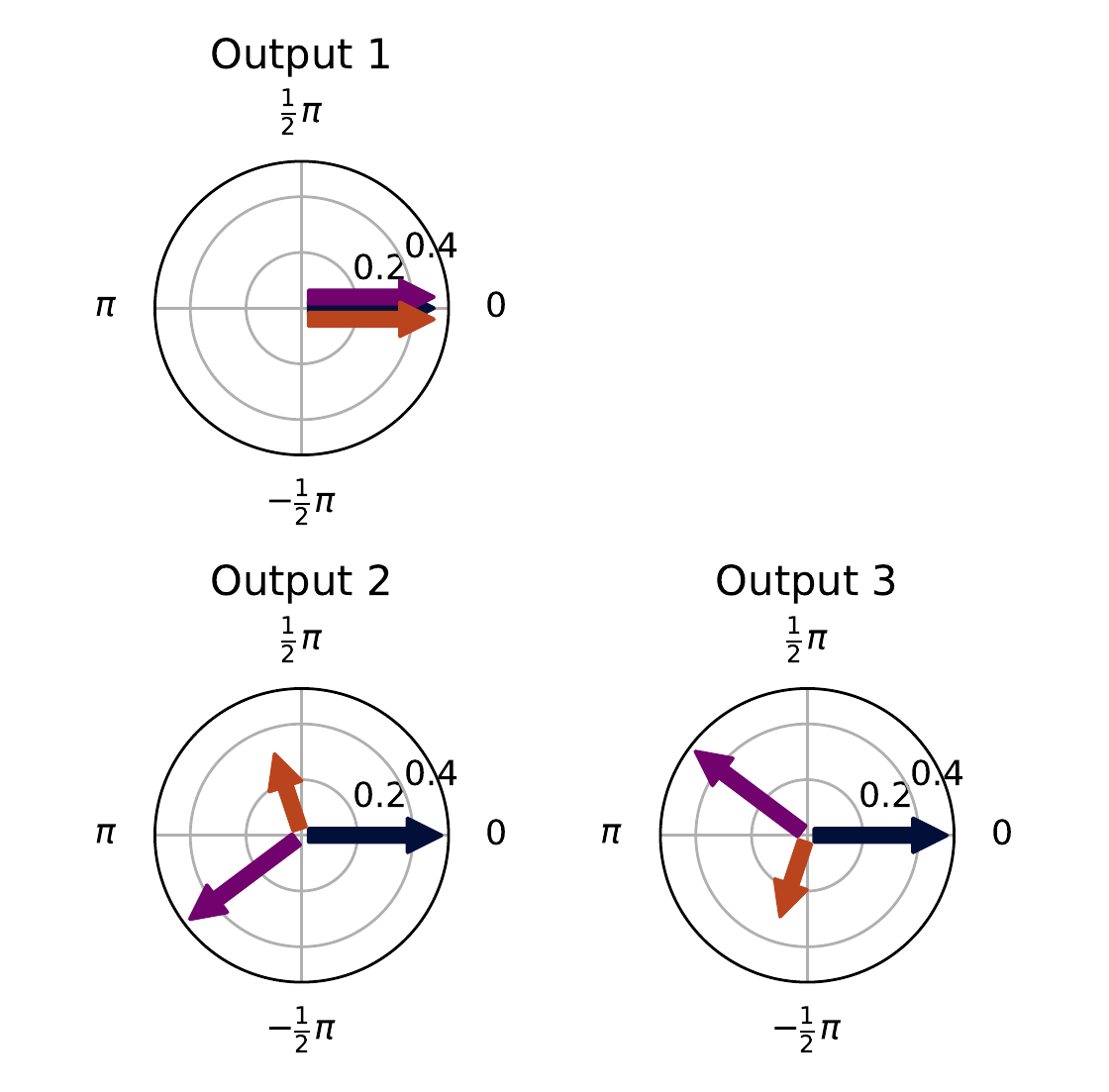}
    \caption{Complex Matrix Plot of the `damaged' X-array type beam combiner with three telescope inputs.}
    \label{Img:CMP_Xarray}
\end{figure}

\begin{figure}
    \centering
    \includegraphics[width=0.75\linewidth]{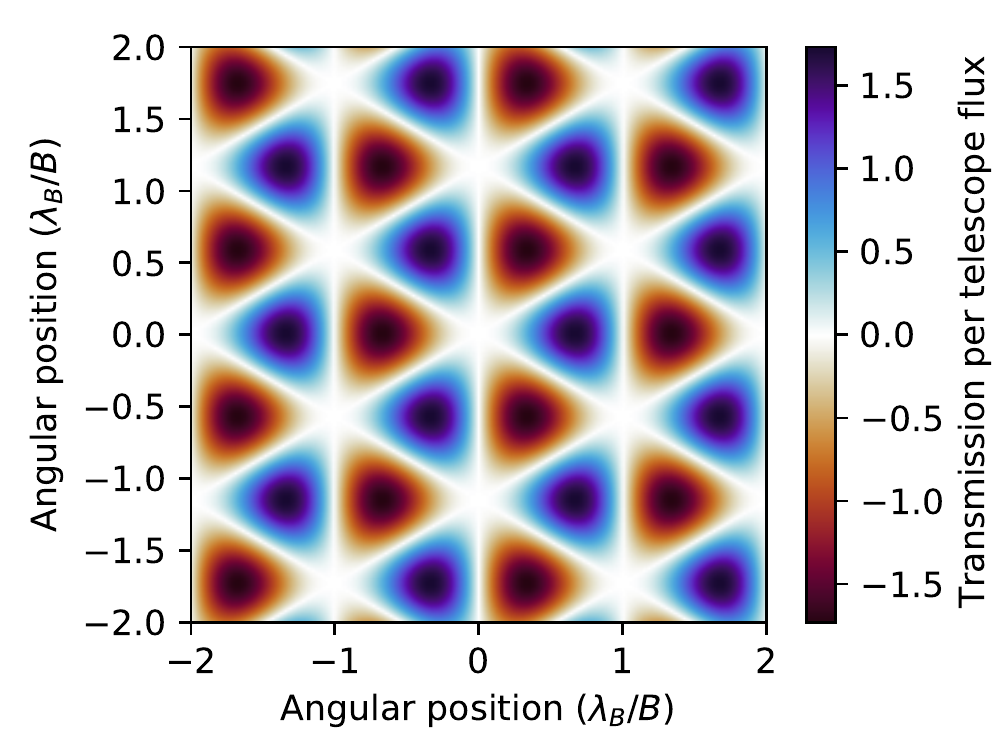}
    \caption{Kernel map of the `damaged' X-array type beam combiner with three telescope inputs.}
    \label{Img:X-array3map}
\end{figure}

Here, we obtain an enantiomorphic pair in outputs two and three as demonstrated in the CMP representation in Figure \ref{Img:CMP_Xarray}. Thus, if a telescope was to fail in this variant of the X-array, the remaining telescopes could move into a triangular position (like in the Kernel-3 nuller of LIFE4) and the beam combiner could still produce a robust observable. Such a map is shown in Figure \ref{Img:X-array3map}.

This map has a maximum transmission of 1.73 telescope fluxes, an efficiency of 58\% compared to a fully functioning three-telescope combiner, or 43\% compared to the undamaged X-array. As before, the shutter will reduce the background in the nulled outputs, this time by a factor of 0.75. This results in an effective S/N of 1.41\red{; 50\% of the original X-array combiner and 66\% of the fully functioning Kernel-3 array.} While this is substantially less than the 100\% efficiency of the X-array with four telescopes, nonetheless this modified combiner would be adequate to continue on the mission in the event of a collector telescope failure.

\section{Conclusion}

In this work, we have provided a practical method to implement a Kernel-5 beam combiner, using a collection of adaptive nullers, spatial filters, beam splitters and phase shifting plates. Adaptive nullers can be used to negate any phase errors induced by imperfections in four of the beam splitting modules, leaving optical errors in the remaining six modules to contribute to errors in the remaining system, including null depth, null stability and kernel sensitivity. These also influence requirements in systematic phase offset errors of the interferometer, as well as RMS fringe tracking errors. 

Taken with a beam splitter reflectance error of $|\Delta R| = 5\%$, and associated phase shift error of $\Delta\phi = 3\degree$, we find that in order to be photon limited and not limited by null fluctuations, we require a fringe tracking error less than 3~nm RMS. Furthermore, in order for the kernels to be appropriately sensitive to planets with a contrast of 1$\times 10^{-7}$ over a bandpass from 4 to 19~\textmu m, we find that the systematic phase error must be less than 0.5~nm. 

We do note, however, that these limits are strongly dominated by the shorter wavelengths, and that at longer wavelengths the requirements lessen substantially. Obtaining high signal in the shorter wavelength regions (around 4~\textmu m) will therefore prove to be harder than at longer wavelengths beyond approximately 8~\textmu m. 

We have also shown a major benefit of the described beam combiner implementation: in introducing a well placed shutter between a coupler of the beam splitter modules, the Kernel-5 combiner can function as a four-telescope combiner. This is a critical advantage if a collecting telescope were to fail or go offline. If these four telescopes were then placed into an X-array configuration, this modified combiner would produce an identical map to the original X-array architecture, albeit with a total throughput penalty of 15\%, split over the two kernel outputs. This is offset by a reduction in the background due to the shutter, resulting in an S/N per telescope equal to 80\% of the fully functional X-array, \red{or an S/N reduction of 37\% compared to the original array}. A further telescope could also be removed with the addition of a second shutter, leading to a Kernel-3 type map with a relative S/N 74\% of the equivalent Kernel-3 beam combiner. Finally, we note that the beam combiner of the X-array itself could be designed in a similar way, and providing the same benefits as the Kernel-5 nuller. If one of the X-array telescopes were to fail, a Kernel-3 type map could be created with an efficiency of 58\% and relative S/N 66\% compared to the Kernel-3 combiner - lower than the equivalent Kernel-5 design, but nonetheless adequate to continue scientific observations.

The next step forward would be to investigate physically constructing such a beam combiner in a laboratory, to test the assumptions about errors and uncertainties in this paper. Furthermore, a more detailed study at the opto-mechanics of injection into a beam combiner like the one described would need to be addressed, for example how four telescopes in a rectangular formation could inject into the combiner designed for five in a pentagonal formation.

The advantage of telescope redundancy, along with the sensitivity advantages as discussed in LIFE4, further adds credence to the Kernel-5 beam combiner, with five telescopes in a pentagonal configuration, as the ideal architecture for the LIFE mission. We therefore suggest that future studies consider adopting this architecture in their analysis of future science and technological requirements for space-based mid-infrared nulling interferometry.

\section*{Acknowledgements}
We acknowledge and celebrate the traditional custodians of the land on which the Australian National University is based, the Ngunnawal and Ngambri peoples, and pay our respects to elders past and present. This research was supported by the ANU Futures scheme and by the Australian Government through the Australian Research Council's Discovery Projects funding scheme (project DP200102383) and the Australian Government Research Training Program. This project has received funding from the European Research Council (ERC) under the European Union’s Horizon 2020 research and innovation programme (Grant agreement No. 866070). We also thank members of the LIFE team for their constructive feedback. Data is available upon request to the author.



\bibliographystyle{aa}
\bibliography{bibliography} 



\appendix

\section{Tuning the null depth}
\label{app:null_tuning}
\red{
As mentioned in the main text, we can use the alignment procedure intermittently during observations to correct for alignment drifts. In this appendix, we show that the integration time required to correct this is small enough to realistically calibrate the null.

First, we assume that for a generic nuller, the stellar flux is much larger than the background.  We first derive the integration time needed to correct an alignment drift of phase for a single nulled output. To correct for this, we assume that we modulate the adaptive nuller by an amplitude of $\pm\Delta \phi$ such that the intensity in the nulled outputs is significantly higher than the background. In this regime, far away from the null bottom, drifts in phase vary quadratically with the intensity ($\phi \propto \sqrt{I}$).

Let the background photon rate per telescope (which can include stellar leakage) be $b$, and the stellar flux rate per telescope be $s$. The nulled output photon rate is then
\begin{equation}
    n = \frac{1}{2}s\Delta\phi^2,
\end{equation}
with total number of photons
\begin{equation}
    N = \frac{1}{2}sT\Delta\phi^2.
\end{equation}

For a simplistic requirement on the uncertainty in $\Delta \phi$, $\sigma$, we require that 
\begin{equation}
    \frac{1}{2}s\sigma^2 << b \Longrightarrow \sigma^2 << \frac{2b}{s}.
\end{equation}
We note that there is a more complex question regarding kernel nulling calibration, which is beyond the scope of this discussion.

We can derive this uncertainty as follows:
\begin{align}
    \frac{\Delta(\Delta\phi)}{\Delta N} &= \sqrt{\frac{1}{2TNs}}\\
   \Delta(\Delta\phi) = \sigma &= \sqrt{\frac{1}{2Ts}},
\end{align}
and so the requirement on $\sigma$ becomes
\begin{equation}
    \frac{1}{2Ts} << \frac{2b}{s} \Longrightarrow T >> \frac{1}{4b}.
\end{equation}

The relevant background rate is the zodiacal light in a single channel, which is about 1 photon per second in a 1\% bandpass at 4~\textmu m. So, as long as the phase offset between the fringe tracker and nuller does not drift on a timescale shorter than 1~s, there is no problem repeating this calibration. This is also short enough to not greatly impact the amount of integration time spent on science data.

The same derivation applies to drifts in amplitude and for each of the outputs, and so with approximately 8 of these calibrations, we need about 8~s of callibration for each science observing block.
}

\section{Derivation of the S/N metric}
\label{app:SNR_metric}
\red{
In this appendix, we give a brief derivation of the S/N equations introduced in the text. Let $f$ be the flux of the planet and $m_i$ be the transmission of the combiner for the $i$th output, kernel or map (henceforth output). Let the noise of this output be $\sigma_i$. The measured signal for each output is then
\begin{align}
    s_i &= m_i \times f.
\end{align}
An estimator for the flux and uncertainty in the flux can be derived:
\begin{align}
    \hat{f} &= \frac{s_i}{m_i} &
    \hat\sigma(\hat{f}) &= \frac{\sigma_i}{m_i}.
\end{align}
The flux-normalised signal to noise ratio, $S/N$, for each output can be written as
\begin{align}
    S/N_i &= \frac{1}{\hat{f}}\frac{\hat{f}}{\hat\sigma(\hat{f})} = \frac{m_i}{\sigma_i}.
\end{align}

Now, the inverse variance weighted average flux, $f_w$, can be calculated with
\begin{align}
    \hat{f_w} &= \frac{\sum_i\frac{\hat{f}}{\hat\sigma(\hat{f})^2}}{\sum_i\frac{1}{\hat\sigma(\hat{f})^2}} = \frac{\sum_i\frac{m_is_i}{\sigma_i^2}}{\sum_i\frac{m_i^2}{\sigma_i^2}},
\end{align}
with a variance of
\begin{align}
    \hat{\sigma}({f_w})^2 &= \frac{1}{\sum_i\frac{1}{\hat\sigma(\hat{f})^2}}
     = \frac{1}{\sum_i\frac{m_i^2}{\sigma_i^2}}.
\end{align}
The weighted flux-normalised S/N is then
\begin{align}
    S/N_{total} &= \frac{1}{\hat{f_w}}\frac{\hat{f_w}}{\hat{\sigma}({f_w})} = \sqrt{\sum_i\frac{m_i^2}{\sigma_i^2}}
    = \sqrt{\sum_i(S/N_i)^2}.
\end{align}
}



\label{lastpage}
\end{document}